\documentclass[12pt,draftclsnofoot,onecolumn]{IEEEtran}
\renewcommand{\baselinestretch}{1.6}%1.55
\ifodd 0

\else

\fi

\ifodd 0

\else

\fi

\newcommand{\mc}[1]{\mathcal{#1}}
\newcommand{\bs}[1]{\boldsymbol{#1}}

\newcommand{\fa}[2]{\, \forall \, #1 \in \mathcal{#2}}

\ifx\UseOption\undefined
\def\UseOption{opt1} %% choose opt1 or opt2
\fi

\usepackage{amssymb,amsmath,color,graphicx}
\usepackage{subfigure}
\usepackage{cite}
\usepackage{verbatim}
\newcommand{\ra}[1]{\renewcommand{\arraystretch}{#1}}
\usepackage{algorithmic}
\usepackage[ruled,linesnumbered,vlined]{algorithm2e}
\SetKwRepeat{Do}{do}{while} %
\usepackage{url}
\usepackage{graphicx,epstopdf}

\DeclareMathOperator*{\argmax}{arg\,max}

\newtheorem{theorem}{Theorem}

\newtheorem{problem}[theorem]{Problem}

\newtheorem{mydef}{Definition}
\newtheorem{mygame}{Game}
\newtheorem{myprp}{Proposition}
\newtheorem{mythm}{Theorem}
\newtheorem{mylem}{Lemma}

\newtheorem{myobs}{Observation}
\usepackage{bm}

\usepackage{multirow}
\usepackage{mathtools}
\begin{document}
	%%%%%%%%%%%%%%%%%%%%%%%%%%%%%%%%%%%%%%%%%%%%%%%%%%%%%%%%%%%%%%%%%%%

\title{\Large{Multimedia Crowdsourcing with Bounded Rationality: \\ A Cognitive Hierarchy Perspective}}

\author{\small{\IEEEauthorblockN{Qi Shao, 
		Man Hon Cheung, 
		and Jianwei Huang\\}
	\IEEEauthorblockA{E-mails: sq017@ie.cuhk.edu.hk, mhcheung@ie.cuhk.edu.hk, jwhuang@ie.cuhk.edu.hk}
}
		\thanks{Qi Shao, Man Hon Cheung, and Jianwei Huang are with the Network Communications and Economics Lab, Department of Information Engineering, the Chinese University of Hong Kong, Shatin, N.T., Hong Kong, China. J. Huang is also with the School of Science and Engineering, the Chinese University of Hong Kong, Shenzhen. This work is supported by the General Research Fund CUHK 14219016 from Hong Kong UGC, and the Presidential Fund from the Chinese University of Hong Kong, Shenzhen. }
	\thanks{Part of this paper was presented in \emph{GLOBECOM'18} \cite{shao_gc18}. 
		%1) Multimedia crowdsourcing: Consideration of quality requirements for multimedia applications. 
		%2) More general model: 2 tasks -> M tasks
		%3) New sim results
		The major differences between the journal and conference versions are:  % as follows
		a) \emph{Multimedia crowdsourcing background}: In \cite{shao_gc18}, we considered a general crowdsourcing platform and studied only the rewards for the tasks. In this journal submission, we further consider the quality requirement together with the reward for each task, which is important for multimedia applications.
		b) \emph{A more general system model}: We extend our crowdsourcing model from two tasks to multiple tasks, which involves solving a challenging mixed integer non-linear programming problem.
		c) \emph{Performance evaluation}: In this journal submission, we have some new simulation results. We demonstrate the impacts of additional parameters that were not considered in \cite{shao_gc18}. %The performance evaluation of the random mobility case is included. %, such as the probability of meeting Wi-Fi,
		To summarize, the overall difference between the journal and conference versions exceeds 60\%.	}
}
%\author{\authorblockN{Qi Shao, Man Hon Cheung, and Jianwei Huang\\}
%		\authorblockA{E-mails: sq017@ie.cuhk.edu.hk, mhcheung@ie.cuhk.edu.hk, jwhuang@ie.cuhk.edu.hk}

\maketitle

% To remove the p.1 in the abstract page
\thispagestyle{empty}

\vspace{-2.2cm}
\begin{abstract}
	%Multimedia crowdsourcing possesses great benefits to exploit the sensing capabilities of mobile devices, while it faces new challenges especially when encountering quality requirements of the multimedia contents. 
	In \emph{multimedia crowdsourcing}, the \emph{requester}'s  quality requirements and reward decisions will affect the \emph{workers'} task selection strategies and the quality of their multimedia contributions.  
	%However, most existing studies on crowdsourcing incentive mechanism design either ignored the quality requirements or simply assumed that the workers are \emph{fully rational}.
	In this paper, we present a first study on how the workers' \emph{bounded cognitive rationality} interacts with and affects the decisions and performance of a multimedia crowdsourcing system.  
	Specifically, we consider a two-stage model, where a requester first determines the reward and the quality requirement for each task, and the workers select the tasks to accomplish accordingly. 
	First, we consider the benchmark case where users are fully rational, and derive the requester's optimal rewards and quality requirements for the tasks. 
	Next, we focus on the more practical bounded rational case by modeling the workers' task selection behaviors using the \emph{cognitive hierarchy} theory. 
	%Each level of worker assumes that his strategy is the most sophisticated and we progressively define the strategic categories: \emph{level-0} workers randomize; and \emph{level-$k$} thinkers best respond, assuming that other workers are distributed from \emph{level-0} to \emph{level-$(k-1)$}. 
	Comparing with the fully rational benchmark, we show that the requester can increase her profit by taking advantage of the workers' bounded cognitive rationality, especially when the workers' population is large or the workers' average cognitive level is low. When the workers' average cognitive level is very high, however, the equilibrium under the practical bounded rational model converges to that under the benchmark fully rational model. It is because workers at different levels make decisions sequentially and high cognitive level workers can accurately predict other users' strategies.
	Under both the fully and bounded rational models, we show that if workers are heterogeneous but one type of workers (either the high or the low quality) dominates the platform, the requester cannot make a higher profit by setting different quality requirements for different tasks.
\end{abstract}

\vspace{-0.1cm}


\normalsize

\newpage
\setcounter{page}{1}
\renewcommand{\baselinestretch}{1.55}
%==============================================================================
\section{Introduction} \label{sec:intro}
%==============================================================================	

%%%%%%%%%%%%%%%%%%%%%%%%%%%%%%%%%%%%%%%%%%%%%%
\subsection{Motivations}
\emph{Mobile Crowdsourcing} \cite{brabham2008crowdsourcing} is a fast-growing outsourcing paradigm that exploits the sensing capabilities of mobile devices from a large group of workers. 
With the advanced functionalities and sensors (e.g., high-resolution cameras, microphones, and GPS) of mobile devices, we can reveal the full potential of crowdsourcing, which may revolutionize many areas in our daily life, such as health care, transportation, and multimedia consumption. 
\emph{Multimedia crowdsourcing} \cite{maharjan2016optimal}, which is a special kind of mobile crowdsourcing for images, audios, and videos, possesses great benefits in areas such as online education, social networking, and environment monitoring. For instance, multimedia crowdsourcing for traffic monitoring can reveal more details than non-multimedia measurements. When a traffic accident occurs, the real-time pictures and videos can help rescuers better estimate the number of people trapped in a particular area more accurately in advance.

Today's multimedia crowdsourcing platforms (e.g., Foap \cite{Foap}) make it possible to hire workers worldwide to accomplish a common task.
In such a system, a \emph{requester} is usually responsible for publishing tasks together with the corresponding rewards and quality requirements, so that the \emph{workers} can claim and complete the tasks to earn these rewards.
For instance, some travel platforms (e.g., Ctrip \cite{ctrip} and KLOOK \cite{klook}) would like to hire some individuals to collect photos or write traveling notes at some scenic spots, aiming at enriching their photo bases and enhancing their reputation. 
Those photographers who produce photos with good resolution and location choices will be rewarded with some coupons or even cash \cite{ctripcollect}.
To achieve her desirable outcome, a requester needs to properly design an \emph{incentive mechanism} on how much reward to give, how to set the quality requirement, and how to share the reward.

The trend of multimedia crowdsourcing attracts more mobile workers to produce and distribute multimedia information and contents.
However, due to the dynamic and distributed characteristics of the multimedia sources (e.g., audios, videos, and graphics), multimedia crowdsourcing imposes more stringent \emph{quality} requirements than usual mobile crowdsensing applications, hence poses new challenges for the incentive mechanism design.
For example, Ctrip prefers photos and travel notes with high quality (i.e., carefully designed photos and elaborated written notes) and aims to recruit professional photographers to upload high quality pictures. To achieve this, Ctrip publishes some photo-taking tasks with high rewards and high quality requirements to attract the high quality workers to complete these particular tasks \cite{ctrip_prof}. 
If the requester sets the requirement too high, however, there will be fewer eligible workers, who can complete these tasks, and it might negatively affect the eventual goal of the requester.  
Thereby, crowdsourcing based multimedia applications involve a complex tradeoff among the rewards, quality requirements, and the workers' participation. However, most existing studies (e.g., \cite{zhao2016budget, zhou2017truthful, han2016truthful, wang2018strategic}) only focus on the reward design without considering the quality requirements.

Furthermore, regarding the workers' participation, existing literature on incentive mechanism designs in crowdsourcing (including multimedia crowdsourcing) usually consider the \emph{fully rational} workers (e.g., \cite{ji2017online, jin_ci17, yang2017designing, song2017quality}), who have the same \emph{infinite} cognitive levels of reasoning other workers' decisions when choosing their own. As a result, the workers' interactions in terms of the task selections are formulated as a \emph{non-cooperative game}, with the \emph{Nash equilibrium} (NE) being the most widely used solution concept \cite{ben1997rationality}.
However, extensive experimental studies in psychology have shown that people often have \emph{cognitive limits} when making their reasoning decisions (e.g., \cite{camerer2004cognitive, crawford2013structural, hossain2013markets}). 
Hence it is more accurate to consider the \emph{bounded rational} workers, who have \emph{heterogeneous} and \emph{finite} cognitive levels of reasoning about the others' decisions.  The cognitive heterogeneity of the workers exerts a great impact on their data collecting decisions, and eventually influences the requester's multimedia information and quality. 
This motivates us to consider the workers’ cognitive capabilities in the modeling of multimedia crowdsourcing systems.

One widely adopted theory that mathematically characterizes the players' limited cognitive capacities is the \emph{cognitive hierarchy} (CH) theory \cite{camerer2004cognitive} (which is part of the \emph{behavioral game theory}). 
Under the CH theory, the players, who are categorized into different cognitive \emph{levels}, reason in a progressive manner to achieve the \emph{cognitive hierarchy equilibrium}. 
More specifically, the \emph{level-0} (lowest cognitive level) players select their strategies \emph{randomly} without predicting the choices of the other players. A \emph{level-1} player, who is one cognitive level higher than the \emph{level-0} player, makes his choice by predicting other users' choices by assuming that all the other players are \emph{level-0} players. In general, a \emph{level-$k$} player assumes that the other players are distributed according to a normalized Poisson distribution \cite{camerer2004cognitive} from \emph{level-0} to \emph{level-$(k-1)$} in the population, and makes his decision by anticipating the other players' decisions accordingly. 
This means that a \emph{level-$k$} player can accurately estimate the \emph{relative} proportions of all the lower level players, but ignores the fact that there can be other players at the same or higher levels than his.
We will give the more detailed mathematical characterization of the CH theory in Section \ref{sec:CH}. 
By analyzing the workers' behaviors from \emph{level-0} to \emph{level-$\infty$} progressively, we can compute the total number of workers selecting each task.
The CH theory has been very successful at explaining deviations from the NE for a wide range of applications, such as marketing \cite{hossain2013markets}, business \cite{cui2017cognitive}, and network science \cite{choi2012cognitive}.

In this paper, we aim to answer the following key questions:

\begin{itemize}
	\item How would the bounded rational workers select the multimedia sensing tasks?
	\item How would the cognitive heterogeneity of the workers influence the requester's profit-maximizing task parameter choices (i.e., reward and quality requirement)?
	\item What is the relationship between the bounded rational model and the fully rational benchmark used in the literature? 
\end{itemize}

%	\vspace{-13pt}
%%%%%%%%%%%%%%%%%%%%%%%%%%%%%%%%%%%%%%%%%%%%%%
\subsection{Contributions}
%	\vspace{-13pt}	

As the first paper on modeling the workers' limited cognitive reasoning capability in multimedia crowdsourcing, we consider a simple model where a single requester recruits some workers to complete multiple tasks. Different tasks involve different completion costs, and generate different revenues for a requester, if completed with different qualities.
We derive the requester's optimal rewards and quality requirements, as well as workers' task selection equilibrium in both the benchmark \emph{fully rational} (FR) model and the \emph{bounded rational} (BR) model. 

We summarize the key results and contributions of this paper as follows:
\begin{itemize}

	\item \emph{Novel crowdsourcing model with the CH theory}: To the best of our knowledge, this is the first paper that applies the CH theory in multimedia crowdsourcing systems, which integrates the quality requirements of tasks and the cognitive heterogeneity of workers.

	\item \emph{Optimal and near-optimal solutions to the FR benchmark}: Under the FR model, we can compute the requester's \emph{optimal} task reward and quality settings and the workers' corresponding NE when the workers have homogeneous quality capabilities. For the case of heterogeneous worker quality capabilities, we propose a low-complexity heuristic algorithm to compute the near-optimal solutions of a mixed integer programming problem.

	\item \emph{Theoretical and numerical analysis in the BR model}: Under the BR model, we show that the task selection equilibrium converges to the FR benchmark in the limiting case when the workers' average cognitive level is high enough. Such an equivalence is not generally true in other systems, and we prove the result by exploiting the special structure of our problem. 
	Under the more realistic case of a finite level of average cognitive level, however, the requester can increase her profit (compared with the FR case) by reducing the reward and taking advantage of the workers with low reasoning capabilities, especially when the workers' population is large or the workers' average cognitive level is low.

	\item \emph{Exploiting workers' heterogeneity in quality}: With heterogeneous workers, in general, the requester can increase her profit by choosing different quality requirements for different tasks (comparing with the single quality requirement for all tasks) under both the FR and BR models. When a single (i.e., high or low) quality capability workers dominate the platform, however, the benefit of quality requirement differentiation disappears, because  the rest of the workers (with a different quality capability than the majority workers) are not enough to make a difference.

\end{itemize}

%\subsection{Paper Organization}
The rest of the paper is organized as follows.
We first review the related literature in Section \ref{sec:literature} and describe our two-stage system model in Section \ref{sec:model}. 
Then in Section \ref{secNEA}, we derive requester's near-optimal task settings and workers' task selections in the FR case.
Next, we introduce the BR model and compare it with the FR model in Section \ref{sec:CH}.
Finally, we demonstrate the performance of these two models through numerical examples in Section \ref{sec:comparison}, and conclude the paper in Section \ref{sec:concl}.

%	\vspace{-10pt}
%%%%%%%%%%%%%%%%%%%%%%%%%%%%%%%%%%%%%%%%%%%%%%
\section{Literature Review} \label{sec:literature} %Literature Review

Although there have been many studies on the incentive mechanism design of crowdsourcing systems, a lot of them (e.g., \cite{zhao2016budget, zhou2017truthful, han2016truthful, wang2018strategic, ji2017online, jin_ci17}) do not consider quality requirements, so they are not applicable to the multimedia crowdsourcing applications.

Several studies consider the incentive mechanism design in the context of multimedia crowdsourcing by jointly considering the quality requirements and rewards (e.g., \cite{maharjan2016optimal, yang2017designing, song2017quality, jin2018thanos}). For example, Maharjan \emph{et al.} \cite{maharjan2016optimal} propose a cloud-enabled multimedia crowdsourcing framework, where workers' optimal service durations depend on the rewards and quality capabilities. Yang \emph{et al.} \cite{yang2017designing} jointly consider quality estimation and monetary incentive, and propose a quality-based truth estimation for crowdsensing. Song \emph{et al.} \cite{song2017quality} and Jin \emph{et al.} \cite{jin2018thanos} develope incentive mechanisms considering workers' quality of information. However, the crowdsourcing workers in these papers are all assumed to be fully rational.

Several recent works in crowdsourcing start to challenge this widely used assumption of full rationality (e.g., \cite{karaliopoulos2016factoring, natalicchio2017innovation, peng2016data}). 
For example, Karaliopoulos \emph{et al.} \cite{karaliopoulos2016factoring} question the fully rational assumption in crowdsourcing systems with the support of empirical data, but they only conduct a comparative study of different behavioral theories.
Natalicchio \emph{et al.} \cite{natalicchio2017innovation} declare that higher reasoning level workers would contribute to a higher quality solution, focusing only on a simulation study.
Peng \emph{et al.} \cite{peng2016data} incorporate the bounded rationality of the contributors into an evolutionary competing game formulation. The authors only consider the homogeneous workers and mainly focus on iterative games and the evolutionary equilibrium solution, which requires a long period for the contributors to learn and adapt. All these papers, however, do not take the quality of data into account, which is crucial for the multimedia crowdsourcing applications.

Camerer \emph{et al.} \cite{camerer2004cognitive} propose the CH theory to explain players' behavioral deviation from the traditional equilibrium that has been widely adopted in the area of economics and management.
Abuzainab \emph{et al.} \cite{abuzainab2017cognitive} apply the CH theory in resource allocation problems, which is the first studying incorporating the CH theory in the field of networking. Inspired by this idea, we introduce the CH theory into our multimedia crowdsourcing framework.

As the existing literature on multimedia crowdsourcing (e.g., \cite{maharjan2016optimal, yang2017designing, song2017quality, jin2018thanos}) do not consider the impact of the workers' bounded rationality, this work is the first study that applies the CH theory to address this open research issue. 
We characterize how the workers' average cognitive level and heterogeneous quality capabilities, together with the population size, influence the requester's and workers' decisions in the crowdsourcing system.

	%==============================================================================
	\section{System Model} \label{sec:model}
	%==============================================================================

	In Section \ref{sec:NEsetting}, we provide a high-level discussion of the multimedia crowdsourcing system.
	Then we introduce the decisions of the requester and the workers in two stages:
	the requester determines the rewards and quality requirements in Stage I (Section \ref{sec:NEstage1}), and the workers decide which task to select in Stage II (Section \ref{sec:stage2}).

	\vspace{-3pt}
	
	%%%%%%%%%%%%%%%%%%%%%%%%%%%%%%%%%%%%%%%%%%%%%%
	\subsection{Multimedia Crowdsourcing Setting} \label{sec:NEsetting}
	\begin{figure}[t]
		\centering
		\includegraphics[width=10cm]{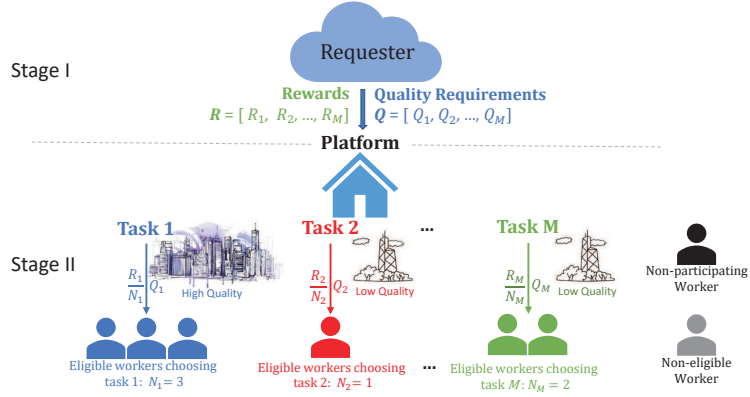}
		\vspace{-12pt}
		\caption{A two-stage multimedia crowdsourcing model. In Stage I, the requester sets rewards $\boldsymbol{R}$ to incentivize the workers to put effort in $M$ tasks, together with quality requirements $\boldsymbol{Q}$. In Stage II, a worker can select to complete one of the tasks or not to participate. Eligible workers (i.e., with equal or higher quality capacities than the requirement) who select the same task will equally share the corresponding reward for that task.}
		\label{fig:mcs}
		\vspace{-12pt}
	\end{figure}
	As shown in Fig. \ref{fig:mcs}, we consider a system where a \emph{requester} announces $M$ tasks on the crowdsourcing platform. 
	The requester associates task $m \in \mc{M} = \{1,\cdots, M\}$ with a reward $R_m \geq 0$ and a quality requirement $Q_m \geq 0$.\footnote{The quality requirements can respond to the pixel requirement of photos and resolution requirements of videos.} For example, the requester publishes some photo collecting tasks at some scenic locations and announces the rewards together with the photo quality requirements. Let $\mathcal{N} = \{1, \cdots, N \}$ be the set of \emph{workers} who are active on this multimedia crowdsourcing platform. 
	We refer to the requester as ``she'' and a worker as ``he'' in the rest of the paper.\footnote{This is just for discussion convenience without any gender preference.}

	As in most multimedia crowdsourcing systems, the requester has a larger market power and makes the decisions before the workers do. Thus, the workers can only treat the requester's decisions as fixed when optimizing their own decisions. Hence we model the interactions in the system as a two-stage game. 
	First, the requester sets the rewards and quality requirements, in order to attract enough workers to complete these tasks with the goal of profit maximization (details in Section \ref{sec:NEstage1}). 
	Next, each worker determines whether to participate and if yes which task to select, in order to maximize his payoff (details in Section \ref{sec:stage2}). We assume that a worker can only complete one task.\footnote{For example, the photo collecting locations are separately far-apart, and a worker cannot visit multiple locations within the time period of interests.}

	%\vspace{-3pt}
	%%%%%%%%%%%%%%%%%%%%%%%%%%%%%%%%%%%%%%%%%%%%%%
	\subsection{Stage I: Requester's Profit Maximization Problem} \label{sec:NEstage1}
	\vspace{-3pt}
	In Stage I, we assume that the requester is fully rational\footnote{With sufficient computational power, it is often reasonable to assume that the requester is fully rational and can choose an optimal strategy \cite{peng2016data}.} and determines the rewards $\boldsymbol{R}=(R_m, \forall m\in\mathcal{M})$ and the quality requirements $\boldsymbol{Q} = (Q_m, \forall m\in\mathcal{M})$ for $M$ tasks to maximize her profit. 
	In Stage II, given rewards $\boldsymbol{R}$ and quality requirements 
	$\boldsymbol{Q}$, the number of workers who are both eligible (with equal or higher quality capabilities than the requirement) and decide to choose task $m \in \mathcal{M}$ is $N_m (\boldsymbol{R}, \boldsymbol{Q})$, which will be derived under the FR and BR models in Section \ref{secNEA} and Section \ref{sec:CH}, respectively. 
	
	Let $U_m(Q_m N_m)$ be the requester's revenue function of task $m \in \mc{M}$, which increases with both the quality threshold $Q_m$ and the number of workers $N_m$ selecting task $m$. 
	First, since the quality requirement determines the minimum quality level of the submitted photos, the contents' quality and hence the requester's revenue increase with $Q_m$.
	Second, when there are more workers collecting data for this task, the requester can have better multimedia contents to select from and hence achieve a higher revenue (e.g., when using these photos to attract future travelers).
	To capture the diminishing marginal return of the additional quality and number of workers, we assume that $U_m(Q_m N_m)$ is an increasing concave function in $(Q_m, N_m)$. 
	
	In Stage I, the requester determines the rewards and quality requirements to maximize her profit (i.e., revenue minus reward) over the $M$ tasks by considering the following problem:

	\begin{problem}
		\underline{\textbf{The Requester's Profit Maximization Problem}}
		\label{prb:1}
		\vspace{-5pt}
		\begin{equation}
		\label{equorg}
		\begin{aligned}
		\text{maximize } & \sum_{m \in \mc{M}} \left( U_m(Q_m N_m (\boldsymbol{R}, \boldsymbol{Q})) -R_m \right)\\
		\text{subject to } & R_m \geq 0, Q_m \geq 0, \forall m \in \mathcal{M},\\
		\text{variables } & \boldsymbol{R}, \boldsymbol{Q}.
		\end{aligned}
		%\vspace{-3pt}
		\end{equation}
	\end{problem}

	 %Because the reward is equally shared and workers are selfish, they will never consume more time or efforts if they meet the minimum quality requirement.  In this case, the requester will provide a higher reward for tasks with higher quality requirements to motivate high quality workers. In Section \ref{sec:homo}, we will consider some concrete function forms of $U_m(Q_m N_m)$. 

	\vspace{-3pt}
	%%%%%%%%%%%%%%%%%%%%%%%%%%%%%%%%%%%%%%%%%%%%%%
	\subsection{Stage II: Workers' Task Selection Problem} \label{sec:stage2}
	In Stage II, for each worker in set $\mathcal{N}$, he needs to decide whether and which task to select.
	We define $q_n$ as worker $n$'s \emph{quality capability} related to his skills and the functionalities of his photographing devices (e.g., smartphones, cameras, and GoPros). The value of $q_n$ is independent of the tasks that he selects. Then we define an eligible task set for worker $n$ as $\mathcal{E}_n(\boldsymbol{Q}) = \{m \in\mathcal{M}: q_n \geq Q_m\}$.

	Let $s_n \in  \{ 0 \}\cup\mathcal{M} $ be worker $n$'s choice (strategy), where ${s_n} = 0$ represents not selecting any task, and ${s_n} = m$ represents selecting task $m \in \mc{M}$.
	When a worker selects (and completes) a task $m \in \mc{M}$, he will incur a cost of $c_m \geq 0$.\footnote{The cost function can include the cost of completing this task, e.g., the cost of fulfilling the requirement and the cost of taking and uploading photos. For example, we can consider a resource consumption model $c_m = c_{m,p} + c_{m,b} \cdot z$, where $c_{m,p}$ is the ``processing cost'' of fulfilling the task requirement, $c_{m,b}$ is the ``bandwidth cost'' related to cellular/Wi-Fi price of uploading the data, and $z$ is the transmitted data size.} We assume that the cost is task dependent but not worker dependent.\footnote{It is practically important to consider the task dependent case. For example, completing simple and small tasks (e.g., taking and uploading photos) incurs a similar amount of cost for different workers. Different types of tasks, however, can lead to very different costs (e.g., taking photos at two different places). In the future, we will consider the more general case of task-dependent and worker-dependent cost. }

	Let $\boldsymbol{s}_{-n} = (s_1, \ldots, s_{n-1}, s_{n+1}, \ldots, s_N)$ be the task selection strategies of all the workers excluding worker $n$.
	Let $\mathcal{N}_m = \{n \in\mathcal{N}: s_n=m, m \in \mathcal{E}_n(\boldsymbol{Q}) \}$ denote the set of workers who are both eligible and decide to select task $ m \in \mathcal{M}$ in Stage II, with a size of $N_m = |\mathcal{N}_m|$. And we have $\sum_{m \in \mc{M}} N_m \leq N$.

	When multiple workers select to complete the same task $m$, the requester will equally divide the reward among all $N_m$ workers.\footnote{We assume that the task is easy to complete and there is no significant difference in terms    of the workers' efforts \cite{zhang_rb12}, hence eligible participating workers have the same possibility of winning the reward, then it is fair to equally split the reward \cite{peng2016data}. Another equivalent implementation is to give the total reward $R_m$ to an arbitrary worker in the set $\mathcal{N}_m$ with an equal probability $\frac{1}{N_m}$. The expected reward of each worker in the set $\mathcal{N}_m$ is $\frac{R_m}{N_m}$.} Hence the payoff of a worker $n$ is 
	\begin{equation}
	\begin{aligned}
	\pi_n (s_n, \boldsymbol{s}_{-n},\boldsymbol{R}, \boldsymbol{Q}) =  \begin{cases}
	0, & \text{if } {s_n} = 0,\\
	\frac{R_m}{N_m} - c_m, &\text{if } {s_n} = m \text{ and } m \in \mathcal{E}_n(\boldsymbol{Q}),\\
	- c_m, &\text{if } {s_n} = m \text{ and } m \notin \mathcal{E}_n(\boldsymbol{Q}).
	\end{cases}
	\end{aligned}
	\label{eq:payoff}
	\end{equation}

	To maximize his payoff, a worker in Stage II will anticipate other worker's strategies, and makes his decision accordingly. In the FR case, workers will have an infinite reasoning capability and can accurately predict the others' behaviors based on (\ref{eq:payoff}). We will formulate the user interactions as a non-cooperative game and derive the NE solution in Section \ref{secNEA}. In the BR case, however, we consider the more realistic assumption that workers have different and limited cognitive levels and may have wrong beliefs about the others' strategies. We derive the corresponding \emph{cognitive hierarchy equilibrium} (CHE) in Section \ref{sec:CH}.
	%Finally, we compare BR model with benchmark FR model in Section \ref{sec:comparison}.

	\section{Fully Rational Model} \label{secNEA}
	\vspace{-1pt}
	%==============================================================================
	In this section, we consider the case when the workers are fully rational. In Section \ref{sec:NEstage2}, we formulate the task selection game in Stage II. Based on this, we compute the sufficient and necessary conditions of the NE for both homogeneous and heterogeneous workers.
	By incorporating such conditions in Stage I, we reformulate the requester's profit optimization problem in Section \ref{sec:bilevel}. Then we present the requester's optimal reward and quality requirement settings and the corresponding NE for homogeneous workers, and 
	propose a low-complexity heuristic algorithm to compute the near-optimal solutions to a mixed integer non-linear programming problem for heterogeneous workers in Section \ref{sec:opthet}.

	%%%%%%%%%%%%%%%%%%%%%%%%%%%%%%%%%%%%%%%%%%%%%%
	\subsection{Workers' Non-cooperative Task Selection in Stage II} \label{sec:NEstage2}
	\vspace{-2pt}
	
	%%%%%%%%%%%%%%%%%%%%%%%%%%%%%%%%%%%%%
	\subsubsection{Non-cooperative Task Selection Game} Workers make their task selection decisions through participating a non-cooperative game, where one worker's payoff depends on the choices of \emph{all} the other workers.
	\begin{mygame}(Workers' Task Selection Game)
		The Stage II's task selection game is a tuple $\Omega = (\mathcal{N} , \boldsymbol{\mathcal{S}},\bm{\Pi})$ defined by:
		\begin{itemize}
			\item Players: The set of workers $\mathcal{N}$.
			\item Strategies: Each worker $n \in \mathcal{N}$ selects a strategy $s_n \in \mathcal{S}_n \triangleq \{ 0 \}\cup\mathcal{M} $. The strategy profile of all workers is $\boldsymbol{s} = (s_n, \forall n \in \mathcal{N})$. The set of feasible strategy profile of all workers is $\boldsymbol{\mathcal{S}} = \times_{n\in\mathcal{N}} \mathcal{S}_n$.
			\item Payoffs: Each worker $n \in \mathcal{N}$ maximizes his payoff as defined in (\ref{eq:payoff}). The vector $\bm{\Pi}(\boldsymbol{s}, \boldsymbol{R}, \boldsymbol{Q}) = (\pi_n(\boldsymbol{s}, \boldsymbol{R}, \boldsymbol{Q}), \forall n\in\mathcal{N})$ contains the payoff functions of all workers.

		\end{itemize}
	\end{mygame}

\subsubsection{Nash Equilibrium}

	For worker $n$, given the rewards $\boldsymbol{R}$, the quality requirements $\boldsymbol{Q}$, and other workers' strategy profile $\bm{s_{-n}}$, he aims to maximize his payoff in (\ref{eq:payoff}):
	
	\begin{problem}
		\underline{\textbf{Worker $n$'s Best Response Strategy Choice Problem}}
		\label{prb:2}
		\vspace{-3pt}
		\begin{equation}
		\label{equpayoff}
		\underset{s_n \in {\mathcal{S}_n}}{\text{maximize }} \pi_n(s_n, \boldsymbol{s}_{-n},\boldsymbol{R}, \boldsymbol{Q}).
		\vspace{-3pt}
		\end{equation}
	\end{problem}

	%After the game definition, we want to derive the equilibrium of the game by computing the best response of each worker. 
	Under a given $\boldsymbol{s}_{-n}$, there can be $\hat{N}_m =|\{k \in \mc{N} \backslash \{n\}: s_k = m, m \in \mathcal{E}_k(\boldsymbol{Q}) \}|$ eligible workers selecting task $m$. After solving Problem \ref{prb:2}, worker $n$'s best response is 
	
	%After the game definition, we want to derive the equilibrium of the game. The first step is to compute the best response of each work, i.e., $s_n^*$.
	%For each worker $n$, given the reward $R_1, R_2$ and the strategy profile $\bm{s_{-n}}$,

	%\vspace{-10pt}
	\begin{equation}
	\begin{aligned}
	s_n^* (\boldsymbol{s}_{-n}, \boldsymbol{R}, \boldsymbol{Q}) = 
	\begin{cases}
	m, & \text{if } m = \underset{i \in \mathcal{E}_n (\boldsymbol{Q})}{\arg \max} \{ \frac{R_i}{\hat{N}_i+1} - c_i\} \text{ and } \frac{R_{m}}{\hat{N}_{m}+1} - c_{m} \geq 0 ,\footnote{ }\\
	0, & \text{otherwise}.
	\end{cases}
	\end{aligned}
	%\vspace{-2pt}
	\end{equation}

	\footnotetext[6]{If multiple tasks satisfy the condition, then worker $n$ will randomly select a task which satisfies the constraint.}
	The fixed point of all the workers' best response choices is the Nash equilibrium (NE), where no worker can improve his payoff by deviating from his task choice unilaterally.
	%We let $(N_1, N_2)$ be the participating NE number of workers achieved under rewards $(R_1, R_2)$.
	
	\begin{mydef}[Nash equilibrium] \label{def:NE}
		Under fixed rewards $\boldsymbol{R}$ and quality requirements $\boldsymbol{Q}$, a strategy profile $\boldsymbol{s}^{NE}$ is an NE of game $\Omega$ if
		%\vspace{-3pt}
		\begin{equation} \label{equ:NE}
		%\begin{aligned}
		\pi_n(s_n^{NE}, \boldsymbol{s}_{-n}^{NE}, \boldsymbol{R}, \boldsymbol{Q}) \geq \pi_n(s_n, \boldsymbol{s}_{-n}^{NE}, \, \boldsymbol{R}, \boldsymbol{Q}),  \quad
		\forall \, s_n \in \, \mathcal{S}_n , \forall  n \in \mathcal{N}.
		%\end{aligned}
		%\vspace{-2pt}
		\end{equation}
	\end{mydef}	
	
	We can write the number of workers selecting a task $m \in \mc{M}$ at the NE $\bs{s}^{NE}$ under rewards $\boldsymbol{R}$ and quality requirement $Q_m$ as
	%\vspace{-3pt}
	\begin{equation}
	N_m^{NE} (\boldsymbol{R}, \bs{Q}) = |\{n \in \mathcal{N}: {s_n^{NE}} = m, m \in \mc{E}_n (\bs{Q}) \} |.
	%\vspace{-3pt}
	\end{equation}

\subsubsection{Analysis} \label{sec:ana2}	
	We will focus on the case when workers have heterogeneous quality capabilities, which includes the case of homogeneous model as a special case.
		
	%\subsection{Heterogeneous Quality Capabilities} \label{sec:hete}
	In the heterogeneous model, some workers possess high quality capabilities and are eligible for completing both high and low quality requirement tasks, while other workers who are less capable can only choose to complete low quality requirement tasks. When workers do not have any differences in terms of capabilities, they are homogeneous. 
	The reason behind this difference is due to workers' different skills and the capabilities of their photographing devices.\footnote{Due to the tractability of the homogeneous model, we can derive the requester's optimal reward and quality requirement. For the heterogeneous model, however, we can only obtain a near-optimal solution, as the corresponding optimization problem is NP-hard. Therefore, we differentiate these two cases and discuss them separately in Section \ref{sec:opthet}.}
	
	More specifically, we assume that the workers have two quality capabilities.\footnote{This is a reasonable simplification for the case where one group of workers carry professional photographic devices and the others rely on smartphones.}
	Without loss of generality, we assume that the $N_H$ workers in the high quality capability set $\mc{N}_H = \{1, \cdots, N_H\}$ have a  quality $q_H$, and the remaining $N-N_H$ workers in the low quality capability set $\mc{N}_L = \{N_H+1, \cdots, N\}$ have a quality $q_L < q_H$.
	
	If there exists a task $m \in \mathcal{M}$ such that $Q_m > q_H$, then none of the workers are eligible for this task and $N_m = 0$. 
	This means that the requester will never choose a $Q_m$ larger than $q_H$.
	
	For the high quality capability workers with $q_H$, they are eligible to complete task $m$ with $0 \leq Q_m \leq q_H$, while low quality capability workers with $q_L$ can only select low quality requirement task $m$ with $0 \leq Q_m \leq q_L$. Hence, we have the quality threshold $q_L$ and then we further define the high quality task set $\mc{M}_H = \{ m \in \mc{M}: q_L < Q_m \leq q_H \}$ only for high quality capability workers, and the low quality task set $\mc{M}_L = \{ m \in \mc{M}: 0 \leq Q_m \leq q_L \}$ which both high and low quality capability workers can select from.
	 %Hence without loss of generality, we restrict the feasible set of $Q_m$ to be $0 \leq Q_m \leq q_H, \forall m \in \mc{M}$. We further define the high quality task set $\mc{M}_H = \{ m \in \mc{M}: 1< Q_m \leq q_H \}$, and the low quality task set $\mc{M}_L = \{ m \in \mc{M}: 0 \leq Q_m \leq 1 \}$.
	
	Note that if we have $N_H = 0$, then all the workers have the same low quality capability, and the heterogeneous setting degenerates to the homogeneous case. In this case, no workers are eligible for the high quality requirement tasks, and the requester will choose $\mc{M}_H = \emptyset$ accordingly.

	Next we characterize the NE strategies of the Workers' Task Selection Game, under the requester's fixed rewards and quality requirements.
	
	\begin{myprp}
		Given rewards $\boldsymbol{R}$ and quality requirements $\bs{Q}$, the numbers\footnote{Usually there are a larger number of workers on the platforms (e.g., Ctrip has 150 million active users), and the decision of a single worker does not have a significant influence on the entire platform (where workers are regarded as non-atomic players \cite{silva2016existence}). Hence, we consider continuous variables $\boldsymbol{N}$ as an approximation of the game formulation in the rest of the paper. For notation simplicity, we use $\boldsymbol{N}^{NE}$ instead of $\boldsymbol{N}^{NE}(\boldsymbol{R}, \boldsymbol{Q})$.} of workers in $\mc{N}_H$ and $\mc{N}_L$ selecting the tasks at the NE, $\boldsymbol{N}_H^{NE} = \left( N_{H,m}^{NE}, \forall m \in \mc{M} \right)$ and $\boldsymbol{N}_L^{NE} = \left( N_{L,m}^{NE}, \forall m \in \mc{M} \right)$, satisfy the following necessary and sufficient conditions: 
		
		\begin{subequations} \label{eq:prphet}
			\begin{align}
			& N_{H,m}^{NE} = \max \left\{0, \frac{R_m}{c_m + \lambda_1} - N_{L,m}^{NE} \right\} , \quad \forall m \in \mathcal{M} , \\
			& \lambda_1 \left( N_H - \sum_{m \in \mc{M}} N_{H,m}^{NE}\right) = 0, \\
			& N_{L,m}^{NE} = 0, \quad \forall m \in \mathcal{M}_H , \\
			& N_{L,m}^{NE} = \max \left\{0, \frac{R_m}{c_m + \lambda_2} - N_{H,m}^{NE} \right\}, \quad \forall m \in \mathcal{M}_L , \\
			& \lambda_2 \left( N_L - \sum_{m \in \mc{M}} N_{L,m}^{NE}\right) = 0, \\
			%& N_{H,m}^{NE}, N_{L,m}^{NE} \geq 0, \quad \forall m \in \mc{M}, \\
			& \lambda_1, \lambda_2 \geq 0, \quad \sum_{m \in \mc{M}} N_{H,m}^{NE} \leq N_H, \quad \sum_{m \in \mc{M}} N_{L,m}^{NE} \leq N_L.
			\end{align}
		\end{subequations}
		\label{prphet}
		\vspace{-3pt}
	\end{myprp}
	
	For a high quality worker who selects task $m$ (i.e., when $N_{H,m}^{NE} = \frac{R_m}{c_m + \lambda_1} - N_{L,m}^{NE} > 0$), constraint (\ref{eq:prphet}a) illustrates that he always receives a payoff of $\frac{R_m}{N_{H,m}^{NE} + N_{L,m}^{NE}} - c_m = \lambda_1 \geq 0$. If such a high quality worker changes to select another task $i$ with $\frac{R_i}{c_i + \lambda_1} - N_{L,i}^{NE} \leq 0$, he will receive a payoff of $\frac{R_i}{N_{H,i}^{NE} + N_{L,i}^{NE}} - c_i \leq \lambda_1$.
	Hence, a high quality worker selecting task $m$ has no incentive to change his strategy.
	If not all the high quality workers participate (i.e., $\sum_{m \in \mc{M}} N_{H,m}^{NE} < N_H$), then we have $\lambda_1 = 0$ from constraint (\ref{eq:prphet}b), which shows that a high quality worker who does not participate will never change to select a task because of the zero payoff of selecting a task. These constraints together show the high quality workers' NE tasks selections.

	For a low quality worker, he is not eligible to select a task with a high quality requirement as shown in constraint (\ref{eq:prphet}c).
	Constraint (\ref{eq:prphet}d) illustrates that a low quality worker selecting task $m \in \mc{M}_L$ has no incentive to change his strategy because he always receives a payoff of $\frac{R_m}{N_{H,m}^{NE} + N_{L,m}^{NE}} - c_m = \lambda_2 \geq 0$.
	If not all the low quality workers participate (i.e., $\sum_{m \in \mc{M}} N_{L,m}^{NE} < N_L$), then we have $\lambda_2 = 0$ from constraint (\ref{eq:prphet}e), which shows that a low quality worker who does not participate will never change to select a task because of the zero payoff of selecting a task. These constraints together show the low quality workers' NE tasks selections. The detailed proof of Proposition \ref{prphet} is given in Appendix \ref{app:prphet}.
	
	Note that the feasible region of variables $\lambda_1$ and $\bs{N}_H^{NE}$ in conditions (\ref{eq:prphet}b) and (\ref{eq:prphet}f) is a union of two convex sets,\footnote{We have $\lambda_1 = 0, \sum_{m \in \mc{M}} N_{H,m}^{NE} \leq N$ or $\lambda_1 > 0, \sum_{m \in \mc{M}} N_{H,m}^{NE} = N$, which is a union of two convex sets. } which is non-convex in general. Due to the same reason, the feasible region of variables $\lambda_2$ and $\bs{N}_L^{NE}$ in conditions (\ref{eq:prphet}e) and (\ref{eq:prphet}f) is non-convex. Given rewards and quality requirements, the feasible set of the corresponding NE is fixed but cannot be computed easily \cite{harker1990finite}. Instead, we directly incorporate these relationships as constraints into the requester's profit maximizing problem and formulate a bilevel optimization problem in Section \ref{sec:bilevel}.

\subsection{Bilevel Optimization} \label{sec:bilevel}
In this subsection, we incorporate the necessary and sufficient conditions (\ref{eq:prphet}a)-(\ref{eq:prphet}f) into Problem \ref{prb:1} in Stage I and form a bilevel optimization problem \cite{harker1990finite}. This enables the requester to compute his optimal choices of rewards and quality requirements in Stage I in (\ref{equorg}) by considering the worker's selections in Stage II.
\begin{problem}
	\underline{\textbf{Requester's Bilevel Profit Maximization Problem}}
	\label{prb:3}
	\vspace{-5pt}
	\begin{subequations} \label{eq:bihet}
		\begin{align}
		{\text{maximize }} &  \sum_{m \in \mc{M}} ( U_m (Q_m ( N_{H,m}^{NE} + N_{L,m}^{NE} ) ) - R_m )\\
		\text{subject to } 
		& R_m \geq 0, \quad \forall m \in \mathcal{M},\\
		& q_L <  Q_m \leq q_H, \forall m \in \mathcal{M}_H, \qquad  0 \leq Q_m \leq q_L, \forall m \in \mathcal{M}_L, \\
		& (\ref{eq:prphet}a)-(\ref{eq:prphet}f), \nonumber \\
		\text{variables } & \boldsymbol{R} , \boldsymbol{Q}, \boldsymbol{N}_H^{NE}, \boldsymbol{N}_L^{NE}, \lambda_1, \lambda_2, \mc{M}_H, \mc{M}_L.
		\end{align}
		%	\vspace{-3pt}
	\end{subequations}
\end{problem}

	Due to the non-convex feasible region of the variables in (\ref{eq:prphet}a)-(\ref{eq:prphet}f), Problem \ref{prb:3} is \emph{non-convex} and is difficult to solve. Nevertheless, we can exploit the special structure of the crowdsourcing problem and simplify the problem formulation.\footnote{We can show that the feasible region is a union of two convex sets. We then compute the optimal solution that achieves the maximum profit on the two convex sets separately. Then we compare these results and derive the optimal rewards and quality requirements.}
	In the Proposition \ref{prphetopt}, we introduce an alternative characterization of the constraints in Problem \ref{prb:3} to simply the problem.

	\begin{myprp}
		The rewards $\boldsymbol{R}^{NE*}$ and the quality requirements $\boldsymbol{Q}^{NE*}$ in Stage I are optimal for Problem \ref{prb:3} only if
		\begin{subequations} \label{eq:hetopt}
			\begin{align}
			& Q_m^{NE*} =  q_H, \forall m \in \mathcal{M}_H, \quad Q_m^{NE*} =  q_L, \forall m \in \mathcal{M}_L,\\
			& R_m^{NE*} =  c_m N_m^{NE*}, \quad \forall m \in \mathcal{M}, \\
		    & N_{m}^{NE*} \geq 0, \forall m \in \mc{M}, \quad \sum_{m \in \mc{M}} {N_m^{NE*}} \leq N, \sum_{m \in \mc{M}_H} {N_m^{NE*}} \leq N_H.
			\end{align}
		\end{subequations}
		\label{prphetopt}
		\vspace{-3pt}
	\end{myprp}

	Since the requester knows the workers' quality capabilities, condition (\ref{eq:hetopt}a) indicates that it is optimal for her to differentiate the tasks with quality requirements and directly set the requirement $Q_m^{NE*}$ equal to the workers' quality capabilities. Hence the requester can take advantage of all the workers and fully utilize their quality capabilities.
	Conditions (\ref{eq:hetopt}b) and (\ref{eq:hetopt}c) together determine the optimal rewards, which are just enough to compensate all the task-selecting workers' operation costs. As a result, any worker who selects a task will achieve a zero payoff, regardless of his quality capability and the task that he selects (including the choice of not to participate).\footnote{This is under the assumption that the requester has the complete information about the workers' quality levels and costs. In reality, the requester can set a reward which covers the workers' costs but provides an extremely small payoff to motivate the workers. For example, the rewards are usually less than 0.1 dollars on the crowdsourcing platforms (e.g., Amazon Mechanical Turk \cite{mturk}).} %The number of high quality workers is limited, so we have $\sum_{m \in \mc{M}_H} {N_m^{NE}} \leq N_H$ in (\ref{eq:hetopt}c), which is different from the homogeneous case.
	The proof of Proposition \ref{prphetopt} is given in Appendix \ref{app:prphetopt}.

Proposition \ref{prphetopt} shows the \emph{necessary} conditions of the requester's optimal rewards and quality requirements in Stage I. We note that if conditions (\ref{eq:hetopt}a)-(\ref{eq:hetopt}c) hold, then all the constraints in Problem \ref{prb:3} are satisfied. 
Thus, we can substitute $N_{H,m} + N_{L,m} = N_m$ and $R_m = c_m N_m$ to denote the rewards in (\ref{eq:bihet}a). We can replace the quality requirement value with $Q_m$ in (\ref{eq:bihet}a) with (\ref{eq:hetopt}a). Equivalently, we can define $y_m \in \{0,1\}$ to represent the quality requirement of task $m \in \mc{M}$. When $y_m = 1$, task $m$ has a high quality requirement  $Q_m = q_H$; when $y_m = 0$, task $m$ has a low quality requirement $Q_m = q_L$. This notation also allows us to represent the high quality task set as $\mc{M}_H = \{m \in \mc{M}: y_m = 1\}$. 

Based on the above discussions, we reformulate Problem \ref{prb:3} as follows:
	\begin{problem}
		\underline{\textbf{Requester's Profit Maximization Problem}}
		\label{prb:4}
	\begin{subequations}
	\begin{align}
	{\text{maximize }} & \sum_{m \in \mc{M}} ( U_m (((q_H - q_L) y_m + q_L) N_m) - c_m N_m )\\
	\text{subject to }
	& N_{m} \geq 0, \forall m \in \mc{M}, \quad \sum_{m \in \mc{M}} {N_m} \leq N,\\
	& y_m = \{0, 1\}, \forall m \in \mc{M}, \sum_{m \in \mc{M}} y_m N_m \leq N_H,\\
	\text{variables } & \boldsymbol{N} , \boldsymbol{y}.
	\end{align}
	\label{refhet}
	\vspace{-20pt}
\end{subequations}
	\end{problem}
 
Note that Problem \ref{prb:4} is a mixed integer non-linear programming (MINLP) problem, which cannot be solved in polynomial time in general \cite{burer2012non}. Thus, we propose a low-complexity heuristic algorithm to solve the problem in Section \ref{sec:opthet}.

\subsection{Low-Complexity Heuristic Algorithm to Solve Problem \ref{prb:4}} \label{sec:opthet}

	In this subsection, we aim to solve Problem \ref{prb:4} using the following idea. 
	First, once the set $\mc{M}_H$ is fixed (i.e., $y_m$ is determined, $\forall m \in \mc{M}$), we notice that the problem reduces to a convex problem as we will state in Problem \ref{prb:5} below. Hence, the maximum profit under a fixed $\mc{M}_H$ can be calculated effectively \cite{boyd2004convex}. 
	Second, we propose a low-complexity heuristic algorithm (i.e., Algorithm \ref{alg:m1}) to compute several candidates of the set $\mc{M}_H$ and select the one achieving the highest profit as the near-optimal task set $\mc{M}_H^{N-OPT}$. (The simulation results in Section \ref{sec:alg} show that Algorithm \ref{alg:m1} can achieve a good performance, where the profit is close to that under the optimal settings (achieved by an exhaustive search). Hence, we refer to our task set $\mc{M}_H^{N-OPT}$ as the near-optimal solution.)
	%Third, given the task set $\mc{M}_H^{N-OPT}$, we can solve the convex problem (\ref{refprohet}), and have the near-optimal rewards and quality requirements accordingly.

	First, let us consider the following reduced problem under a fixed high quality task set $\mc{M}_H$:
\begin{problem}
	\underline{\textbf{Requester's Profit Maximization Problem under a Fixed $\mathcal{M}_H$}} 
	\label{prb:5}
\begin{subequations}
\begin{align}
{\text{maximize }} & \Pi(\mc{M}_H, \bs{N}) \triangleq \sum_{m \in \mc{M}_H} ( U_m (q_H N_m ) - c_m N_m ) + \sum_{m \in \mc{M} \backslash \mc{M}_H} ( U_m (q_L N_m ) - c_m N_m )\\
\text{subject to } 
& N_m \geq 0, \forall m \in \mc{M}, \quad \sum_{m \in \mc{M}} {N_m} \leq N, \text{ and } \sum_{m \in \mc{M}_H} {N_m} \leq N_H,\\
\text{variables } & \boldsymbol{N}.
\end{align}
\label{refprohet}
\vspace{-5pt}
\end{subequations}
\end{problem}

	Note that for the case of homogeneous workers, we have $N_H = 0$ and $\mc{M}_H = \emptyset$. Hence we can directly compute the optimal solutions to the convex problem \ref{prb:5} as $\bs{N}^{NE*}$, and have the optimal rewards $\bs{R}^{NE*}$ and quality requirements $\bs{Q}^{NE*}$ from Proposition \ref{prphetopt}.

For the case of heterogeneous workers, we further define the optimal solutions to the convex Problem \ref{prb:5} as $\hat{\boldsymbol{N}}(\mathcal{M}_H)$. Then, the maximum profit under this fixed $\mc{M}_H$ can be calculated by applying $\hat{\boldsymbol{N}}(\mathcal{M}_H)$ into the objective function (\ref{refprohet}a), which we define as $\Pi(\mc{M}_H, \hat{\boldsymbol{N}}(\mathcal{M}_H))$.

Next we will derive the optimal set $\mc{M}_H$ in order to solve Problem \ref{prb:4}. Note that there are totally $2^M$ feasible choices of $\mc{M}_H$, hence the complexity of finding $\mathcal{M}_H$ through the exhaustive search is high. Since the requester in crowdsourcing applications usually needs a fast algorithm in the task assignment process (e.g., real-time tasks such as emergency response \cite{boutsis2014task}), we propose the heuristic Algorithm \ref{alg:m1} with a polynomial time complexity $O(M^2)$ to achieve a suboptimal performance.
Here, we want to clarify that Algorithm \ref{alg:m1} is not to solve Problem 5. Instead, we apply Algorithm \ref{alg:m1} and the solution of convex Problem 5 one after another to find a near-optimal solution of Problem 4.

%balance the complexity with the performance. Then in the following bullets, need to discuss things more concretely, such as the choice of \alpha and the choice of MaxIter. Keep in mind that explaining things clearly is only the first level of technical writing. The more important part is to explain why we do certain things, what are the challenges and difficulties, and what are the contributions and insights.

\renewcommand{\baselinestretch}{1}
	\begin{algorithm} [t]
	 %[You can tune the number]
	\caption{Greedy Randomized Task Set Search Algorithm for Computing Near-Optimal Set $\mathcal{M}_H^{N-OPT}$}  
	\label{alg:m1}  
	
	\KwIn{Utility vector $\boldsymbol{u}$; cost $\boldsymbol{c}$; total number $N$; high quality workers' number $N_H$; task set $\mathcal{M}$; adaptive parameter $\alpha$; number of iterations $MaxIter$.}

	 Initialization: $MaxPro = 0$  \;
	
	% Sort tasks according to the utility: $u_1 \geq u_2 \geq \cdots \geq  u_M$ \;
	\For{$Iter = 1$; $Iter \leq MaxIter$; $Iter = Iter + 1$}{
		
	Set $\mc{M}_H = \emptyset$; Solve for $\hat{\bs{N}}(\mc{M}_H)$ in problem \ref{prb:5}; Calculate $profit = \Pi \left( \mc{M}_H,\hat{\bs{N}}(\mc{M}_H) \right)$ \;
	
	\For{$i =1$; $i \leq M$; $i = i+1$}{
		\ForEach{$j \in \mc{M} \backslash \mc{M}_H$}
		{Set $\mc{M}_H' = \mc{M}_H \bigcup \{j\}$ ; Solve for $\hat{\bs{N}}(\mc{M}_H')$ in problem \ref{prb:5} \;
			Calculate $P_j = \Pi \left( \mc{M}_H',\hat{\bs{N}}(\mc{M}_H') \right)$ \;}

		$P_{h} = \underset{j \in \mc{M} \backslash \mc{M}_H}{\max} P_j$ ; $P_{l} = \underset{j \in \mc{M} \backslash \mc{M}_H}{\min} P_j$ ; $P^* = P_{l} + \alpha (P_{h} - P_{l})$ \;
		
		$\mathcal{J}^* = \{ j^* \in \mc{M} \backslash \mc{M}_H: P_{j^*} \geq P^* \}$ ; Randomly select a task $j^* \in \mathcal{J}^*$ \;
		
		\eIf{$P_{j^*}\geq profit$}{Set $\mc{M}_H = \mc{M}_H \bigcup \{j^*\}$ ; Solve for $\hat{\bs{N}}(\mc{M}_H)$ in problem \ref{prb:5} \;
		 Calculate $profit = \Pi \left( \mc{M}_H,\hat{\bs{N}}(\mc{M}_H) \right)$ \;}{break \;}

	}

\If{$profit > MaxPro$ }{
$MaxPro = profit$ ; $\mathcal{M}_H^{N-OPT} = \mc{M}_H$ \;
}	
}

	%\ENSURE ~~\\ %算法的输出：Output  
	\KwOut{$\mathcal{M}_H^{N-OPT}$.} 
	%\end{algorithmic}  
	\vspace{-1pt}	
\end{algorithm}

\renewcommand{\baselinestretch}{1.55 }

\begin{itemize}

	\item \emph{Initialization:}  
	We initialize the maximum profit $MaxPro$ to be zero (line 1). 
	
	\item \emph{Iteration process:}  
	At each iteration, we determine the task set $\mc{M}_H$ by adding different tasks and compute the corresponding profit. If the task set has a higher profit than that of the previous task sets, then we adopt the new task set $\mc{M}_H$ as the near-optimal set $\mc{M}_H^{N-OPT}$.

	\begin{itemize}
		\item \emph{Preliminary:}  At the beginning of each iteration, we form a new task set $\mc{M}_H = \emptyset$ and compute the corresponding profit $profit$ (line 3).

		\item \emph{Greedy randomized task set search:} When adding a task $j \in \mc{M} \backslash \mathcal{M}_H$ into the task set $\mc{M}_H$ (line 6), we can calculate the profit $P_j$ (line 7). Then, based on the greedy randomized adaptive search technique \cite{resende2010greedy}, we restrict the feasible task candidate set $\mathcal{J}^*$ to be the set that includes all new tasks with profit no smaller than $P^* = P_{l} + \alpha (P_{h} - P_{l})$ (line 8)\footnote{Here $ \alpha \in [0,1]$ is the adaptive rate. Here $\alpha = 1$ corresponds to a greedy algorithm which only adopts a new task which generates the highest profit at each iteration, while $\alpha = 0$ corresponds to a totally random selection process. Since a greedy algorithm cannot escape from a local optimal solution, and a totally randomized process needs more iterations to find a good performance solution \cite{resende2010greedy}, we adopt $\alpha = 0.5$ in our simulations. More discussions regarding the impact of $\alpha$ and $maxIter$ values on the convergence and optimality is given in Section \ref{sec:alg}.}. We then randomly select a task $j^* \in \mathcal{J}^*$ and add task $j^*$ into the task set $\mc{M}_H$ (line 9).
		
		\item \emph{New task set update:} If the new task set $\mc{M}_H$ has a profit higher than that of the previous task set (without task $j^*$) (line 10), then we update the task set (line 11) together with the new profit value (line 12). Otherwise, we terminate this iteration process (line 14). This is because the number of workers with high quality capabilities is limited, we will not always obtain a higher profit by adding a new task into the high quality requirement task set.
	\end{itemize}

	\item 	\emph{Output:} At each iteration, we compare and update the near-optimal task set $\mathcal{M}_H^{N-OPT}$ as well as the maximum profit $MaxPro$ (line 16). This process repeats for $ MaxIter$ times,\footnote{The larger the number of iterations $ MaxIter$, the larger will be the computation time and the better the solution \cite{resende2010greedy}. We adopt $MaxIter = 20 M$ (where $M$ is the total number of tasks) to achieve the best empirical tradeoff between the computation time and the solution performance. The detailed simulation is given in Section \ref{sec:alg}.} when we have enough iterations to have the near-optimal set $\mathcal{M}_H^{N-OPT}$. 
	%	\vspace{-3pt}
\end{itemize}

Third, by applying Algorithm \ref{alg:m1} together with the optimal solutions to Problem \ref{prb:5}, the requester can determine the near-optimal set $\mathcal{M}_H^{N-OPT}$ and the corresponding $\boldsymbol{\hat{N}} (\mathcal{M}_H^{N-OPT})$. We then compute near-optimal rewards $\boldsymbol{R}^{N-OPT}$ and quality requirements $\boldsymbol{Q}^{N-OPT}$ in Stage I based on Proposition \ref{prphetopt} as follows:
\begin{equation}
\begin{aligned}
& Q_m^{N-OPT} =  q_H, \forall m \in \mathcal{M}_H^{N-OPT}, \quad Q_m^{N-OPT} =  q_L, \forall m \in \mathcal{M} \backslash \mathcal{M}_H^{N-OPT},\\
& R_m^{N-OPT} =  c_m \hat{N}_m(\mathcal{M}_H^{N-OPT}), \quad \forall m \in \mathcal{M}.
\end{aligned}
\end{equation}

%In the fully rational model, every participating worker will receive a reward of $\frac{R_m^{N-OPT}}{N_m^{NE}(\boldsymbol{R}^{N-OPT}, \boldsymbol{Q}^{N-OPT})} = c_m$, which equals to his operation cost of selecting task $m$, regardless of the quality requirements and the total number of workers. 
We list the key notations of the FR and BR models in Table \ref{table:not}.

%%%%%%%%%%%%%%%%%%%%%%% ttt %%%%%%%%%%%%%%%%%%%%%%%%%%%%%%%%%%%
\begin{table} [t]
	%\renewcommand{\arraystretch}{1.3}
	\caption{Key Notations} % (a)-(d)
	\label{table:not}
	\vspace{-5pt}
	\centering
	\ra{0.85}
	\begin{tabular}{@{}ccccc@{}} 
		%\begin{tabular}{|c|c|c|}
		\hline
		\multicolumn{3}{c}{\multirow{2}{3cm}{\textbf{Meaning}} } & \textbf{FR Model} & \textbf{BR Model} \\
		
		&&&\textbf{(Section \ref{secNEA})} & \textbf{(Section \ref{sec:CH})} \\
		
		\hline
		
		\multirow{ 4}{*}{Stage I} & \multirow{ 2}{*}{Optimal} & Rewards  & $\boldsymbol{R}^{NE*}$ & $\boldsymbol{R}^{CHE*}$ \\
		\cline{3-5}
		
		&& Quality requirements  & $\boldsymbol{Q}^{NE*}$ & $\boldsymbol{Q}^{CHE*}$ \\
		\cline{2-5}
		
		& \multirow{ 2}{*}{Near-optimal} & Rewards  & $\boldsymbol{R}^{N-OPT}$ & - \\
		\cline{3-5}
		
		&& Quality requirements  & $\boldsymbol{Q}^{N-OPT}$ & - \\

		\hline
		
		\multicolumn{3}{c}{Task Selection Equilibrium in Stage II} & $\boldsymbol{N}^{NE}(\boldsymbol{R}, \boldsymbol{Q})$ & $\boldsymbol{N}^{CHE}(\boldsymbol{R}, \boldsymbol{Q})$ \\

		\hline
		
	\end{tabular}
	%	\vspace{-2pt}
\end{table}

%==============================================================================
\section{Bounded Rational Model} \label{sec:CH}
%==============================================================================
Extensive experimental studies have shown that people have cognitive limits when making their reasoning decisions (e.g., \cite{camerer2004cognitive, crawford2013structural, hossain2013markets}). Instead of making the often unrealistic full rationality assumption, in this section we consider the more practical BR model, where the workers have different reasoning capabilities. We model both the homogeneous and heterogeneous workers' behaviors based on the CH theory in Section \ref{sec:CHtheory}, and theoretically compare the workers' NE and CHE task selections given the same rewards and task requirements in Section \ref{sec:specialt}. Specifically, we prove that the FR model is a limiting case of the BR model when the average cognitive level $\tau$ approaches $\infty$ (i.e., all the users have an infinite reasoning capacity).\footnote{Note that this convergence is not generally true in the CH theory \cite{camerer2004cognitive}. For example, some of the games do not have any NE. Some games have multiple NEs, but the CHE only converges to a particular NE \cite{hossain2013markets}.}

%%%%%%%%%%%%%%%%%%%%%%%%%%%%%%%%%%%%%%%%%%%%%%
\subsection{Workers' Task Selections in Stage II: Cognitive Hierarchy} \label{sec:CHtheory}

To maximize his payoff in (\ref{eq:payoff}), a worker in Stage II consider choosing a task from set $\mc{M}$. If $R_m < c_m$ for a task $m\in \mathcal{M}$, a worker will never select that task as his payoff will be negative (even if he is the only one selecting the task). Hence, for the rest of the paper, we will focus on the non-trivial case of $R_m \geq c_m$ for each task $ m \in \mc{M}$. 

Next, we model the BR case based on the CH theory \cite{camerer2004cognitive}, where workers with different cognitive thinking levels make decisions differently based on different beliefs of the other workers' choices. 
We introduce the basic ideas of the CH theory similar as in  \cite{camerer2004cognitive}:

\begin{itemize}
	\item [1)] The total worker population can be divided into an infinite number of levels, indexed by $k=0,1, \cdots, \infty$. The number of workers at each \emph{level} $k$ follows a Poisson distribution \cite{camerer2004cognitive} $f(k) = \frac{\tau ^ k }{k!} e^{-\tau}$ with a rate $\tau$.
	The value of $\tau$ represents the average cognitive level of the population. For example, a larger $\tau$ may reflect a higher average worker education level.

	\item[2)] A \emph{level-$k$} ($k \geq 1$) worker knows the accurate \emph{relative ratios} among workers at lower rationality levels, $f(h)$ for all $0 \leq h \leq k-1$, and can accurately predict their behaviors, but ignores the existence of other \emph{level-$k$} and higher level workers. 
	More specifically, a \emph{level-$k$} worker's belief about the fraction of \emph{level-$h$} ($0 \leq h \leq k-1 $) is $g_k(h)=\frac{f(h)}{\sum_{j=0}^{k-1}f(j)}$, where the subscript ``$k$'' denotes that such belief is unique to the \emph{level-$k$} worker. Furthermore, a \emph{level-k} worker believes that he is the only one with the highest cognitive capability in the population, i.e., he is the only \emph{level-k} worker in the population, and there are no workers higher than \emph{level-$k$}.\footnote{Suppose that there are 200 \emph{level-0}, 400 \emph{level-1}, and 300 \emph{level-2} workers in the platform. The numbers are chosen for an easy illustration and does not correspond to a particular Poisson distribution. A \emph{level-1} worker believes that all other 899 workers are \emph{level-0}. A \emph{level-2} worker has the belief that $g_2(0) = \frac{200}{200+400} = 0.33$ and $g_2(1) = \frac{400}{200+400} = 0.67$, i.e., there are 299.7 \emph{level-0} workers, and 599.3 \emph{level-1} workers. Here for simplicity we assume that the number of workers can be non-integer. This significantly simplifies the analysis and does not bring much error when the worker population is large enough. } Hence for any $h \geq k$, we have $g_k(h) = 0$. A \emph{level-$k$} worker believes that \emph{level-$h$} workers account for $g_k(h)$ fraction of the population.\footnote{Recall that in the FR model, every worker can correctly anticipate all other workers' decisions in Stage II. In the BR model, however, workers at different levels have different beliefs regarding the workers' population composition and only anticipate the choices of workers at lower levels. }

	\item[3)] Based on the discussions of 1) and 2), we can compute the choices of workers at different levels progressively. 
	Given $R_m \geq c_m$, a \emph{level-0} worker believes that he will get a non-negative payoff no matter which task to select (as he ignores the choices of any other worker in the population), hence he will select an eligible task randomly. A \emph{level-$k$} worker will select a task that leads to the maximum payoff (including the choice of not participating), considering the task selections of all other workers (which are assumed to be at lower levels).
\end{itemize}

Given any arbitrary rewards $\bs{R}$ and quality requirements $\bs{Q}$ in Stage I, we propose a progressive task selection algorithm\footnote{Note that Algorithm \ref{alg:CH} can be applied to both the homogeneous and heterogeneous quality capabilities case. For homogeneous case, the input number of high quality workers should be $N_H = 0$.} (i.e., Algorithm \ref{alg:CH}) motivated by the idea from \cite{camerer2004cognitive}, and compute the CHE $(\boldsymbol{N}^{CHE})$ in Stage II. Notice that the application setting in \cite{camerer2004cognitive} is different from here: the players only decide whether to participate or not in \cite{camerer2004cognitive}, while workers have more task selection choices in our model.

\renewcommand{\baselinestretch}{1}
\begin{algorithm} [t]
%	\footnotesize
	\caption{Task Selection Algorithm based on CH }  
	\label{alg:CH}  
	
	\KwIn{Rewards $\boldsymbol{R}$; quality requirements $\boldsymbol{Q}$; costs $\boldsymbol{c}$; average cognitive level $\tau$; total number of workers $N$; worker number of high quality workers $N_H$; task set $\mathcal{M}$; tolerant error $\epsilon$.} 
	
	Compute $\mc{M}_H = \{ m \in \mc{M} : q_L < Q_m \leq q_H \}$ and $\mc{M}_L = \{ m \in \mc{M} : 0 \leq Q_m \leq q_L \}$ \;

	Set $k = TF = n_m^H = n_m^L = 0$ ; $e^H(0,m) = \frac{1}{|\mc{M}|}, \forall m \in \mathcal{M}$ ; $e^L(0,m) = \frac{1}{|\mc{M}_L|}, \forall m \in \mc{M}_L$\;
	
	\Repeat{$TF > 1 - \epsilon$ }  {
		$f(k) = \frac{\tau ^ k }{k!} e^{-\tau}$ ; $TF = TF + f(k)$ ; $k = k + 1$ \;
		
		\ForEach{$m \in \mathcal{M}$}{
			$n_m^H = n_m^H +  f(k) \cdot e^H(k,m) \cdot N_H $ ; $n_m^L = n_m^L +  f(k) \cdot e^L(k,m) \cdot (N - N_H) $ \;
		
		\eIf{$m \in \mathcal{M}_H$}{$E(k,m)=\frac{R_m}{\frac{n_m^H}{TF}} - c_m$ \; }{$ E(k,m)=\frac{R_m}{\frac{n_m^H + n_m^L}{TF}} - c_m$  \; } }

		$\mathcal{M}_L^* = \left\{ m^* \in \underset{m \in \mc{M}_L}{\argmax}  E(k,m) \right\}$ ; $\mathcal{M}_H^* = \left\{ m^* \in \underset{m \in \mc{M}}{\argmax}  E(k,m) \right\}$ \;

	%	\ForEach{$m \in \mathcal{M}$}{
			\ForEach{$m \in \mathcal{M}$ \textnormal{ and } $i \in \{  H, L \}$}{
			\eIf{$ m \in \mathcal{M}_i^*$ \textnormal{and} $E(k,m) \geq 0$}
			{$e^i(k, m) = \frac{1}{|\mathcal{M}_i^*|} $ \; }{$e^i(k,m) = 0$ \;}
		% }	
		
	}
		
	}

	%\ENSURE ~~\\ %算法的输出：Output  
	\KwOut{CHE Solution: $\boldsymbol{N}^{CHE} = (n_m^H + n_m^L, \forall m \in \mc{M}) $.} 
	%\end{algorithmic}  
	\vspace{-1pt}	
\end{algorithm} 

\renewcommand{\baselinestretch}{1.55}

\begin{itemize}
	\item \emph{Preliminary:} Given the quality requirements, we compute the task set $\mc{M}_H$ and $\mc{M}_L$ with high and low quality requirements respectively (line 1).  
	
	\item \emph{Initialization:}  
	We initialize workers' cognitive level as $k = 0$, the total fractions of workers that have been considered by \emph{level-$k$} workers as $TF = 0$ (line 2). We initialize the number of high and low quality workers selecting task $m$ as $n_m^H =0$ and $n_m^L = 0$ respectively (line 2).
	We initialize the fraction of \emph{level-0} workers with high and low quality capabilities selecting task $m$ as $e^H(0,m) = \frac{1}{|\mc{M}|}$ and $e^L(0,m) = \frac{1}{|\mc{M}_L|}$, respectively (line 2).\footnote{Sometimes, workers in a particular level may find that working on a number of tasks can lead to the same non-negative payoff. In this case, there will be the same fraction of workers in this level choosing each task. Hence, we focus on the fraction and then the expected number of workers in this level selecting these tasks.}
	
	\item \emph{Iteration process:}  
	At iteration $k$, we compute the number of \emph{level-$k$} workers selecting a task $m \in \mc{M}$. Given the task selections of \emph{level-0} to \emph{level-$k$} workers, a \emph{level-$(k+1)$} worker can compute his expected payoff if he selects a task $m\in\mc{M}$.

	\begin{itemize}
		\item \emph{Expected payoffs:} We compute \emph{level-$k$} workers' fraction $f(k)$ (line 4) and the number of workers selecting task $m$ (line 6). Then each \emph{level-$(k+1)$} worker believes that the number of workers selecting a task $m \in \mc{M}_H$ is $\frac{n_m^H}{TF}$, while the number of selecting a task $m \in \mc{M}_L$ is $\frac{n_m^H + n_m^L }{TF}$. Then he computes his expected payoff $E(k,m)$ when task $m$ has a high (or low, respectively) requirement (line 8 or line 10, respectively).
		
		\item \emph{Selecting tasks:} For workers in \emph{level-$(k+1)$} with low (or high, respectively) quality capabilities, they will select the tasks with maximum payoff. In other words, if task $m$ is in the optimal task set and the payoff is non-negative, then there will be $\frac{1}{|\mc{M}_L^*|}$ (or $\frac{1}{|\mc{M}_H^*|}$, respectively) fraction of \emph{level-$(k+1)$} workers choosing task $m$ (line 14); otherwise, none of the \emph{level-$(k+1)$} workers will select task $m$ (line 16).
	\end{itemize}

	\item 	\emph{Output:}  This process repeats until $ TF > 1 - \epsilon$, when we have considered the reasoning of almost\footnote{Although $k$ cannot reach $\infty$ to recruit all the workers, a small $\epsilon$ value (e.g., $10^{-3}$) can guarantee that most (over 99.9\%) workers in the platform are taken into consideration. However, a small $\epsilon$ value also incurs more iterations and computational time.} all the workers. The total number of workers selecting a task $m\in \mathcal{M}$ at the CHE is $N_m^{CHE} = n_m^H + n_m^L$.
	%	\vspace{-3pt}
\end{itemize}

\vspace{-3pt}
%%%%%%%%%%%%%%%%%%%%%%%%%%%%%%%%%%%%%%%%%%%%%%
\subsection{What is the Connection between the FR and BR Models?} \label{sec:specialt}
In this section, we show that the theoretically elegant FR model widely used in the literature is a limiting case of the more practical BR model (as $\tau$ approaches $\infty$) in the context of our crowdsourcing system.\footnote{Camerer \emph{et al.} \cite{camerer2004cognitive} showed that as $\tau \to \infty$, the CHE converges to an NE, if the NE can be reached by finitely many iterated deletions of weakly dominated strategies. However, our crowdsourcing model here does not belong to this category of games.}
This means that the insights derived by the vast incentive mechanism design literature for crowdsourcing are still good approximations of the reality when the workers have sufficiently high reasoning capabilities  (although this is not always true in practice).   

For a fair comparison regarding the workers’ choices between the FR and BR models, we will set the same $\boldsymbol{R}$ and $\boldsymbol{Q}$ in Stage I for both the FR and BR models.
Then we derive and compare the number of workers selecting task $m \in \mathcal{M}$ in Stage II, which we will denote as $N_m^{NE}(\boldsymbol{R}, \boldsymbol{Q})$ and $N_m^{CHE} (\boldsymbol{R}, \boldsymbol{Q})$ in the FR and BR models, respectively.

\subsubsection{Homogeneous Worker Quality Capabilities}
When the workers have the same quality capability, all of them are eligible for all the tasks. 
Since a worker cannot select multiple tasks simultaneously, we have $\sum_{m \in \mathcal{M}} N_m^{NE}(\boldsymbol{R}, \boldsymbol{Q}) \leq N$. We will theoretically discuss the case of $\sum_{m \in \mathcal{M}} N_m^{NE}(\boldsymbol{R}, \boldsymbol{Q}) = N$ in Theorem \ref{thmall}.

\begin{mythm}
	\label{thmall}
	%Given the rewards and quality requirements $\boldsymbol{R}$ and $\boldsymbol{Q}$, and 
	Consider the case where the number of homogeneous workers selecting tasks at the NE satisfies $\sum_{m \in \mathcal{M}} N_m^{NE}(\boldsymbol{R}, \boldsymbol{Q}) = N$. If we choose the same $\boldsymbol{R}$ and $\boldsymbol{Q}$ in the BR case,\footnote{We have generalized this result comparing with Theorem 2 in \cite{shao_gc18}, where we restricted the rewards to be $\bs{R} = \bs{R}^{NE*}$.} then we have the following results: (a) every worker will participate by selecting one of the $M$ tasks in the BR case (i.e., $s_n \neq 0, \forall n \in \mc{N}$); (b) as $\tau$ approaches $\infty$, the CHE $N_m^{CHE}(\boldsymbol{R}, \boldsymbol{Q})$ converges to the NE $N_m^{NE}(\boldsymbol{R}, \boldsymbol{Q})$, $\forall m \in \mathcal{M}$. 
\end{mythm}

Theorem \ref{thmall} shows that if the rewards are high enough for all the workers to participate under the FR case, then the same rewards will also motivate all the BR workers to participate.
The detailed proof is given in Appendix \ref{app:thmall}.

\subsubsection{Heterogeneous Worker Quality Capabilities}
When the workers have heterogeneous quality capabilities, high quality workers can select all the tasks, while the low quality ones can only select the tasks with low quality requirements.
Now we discuss the workers' behaviors in Theorem \ref{thmhigh}.

\begin{mythm}
	\label{thmhigh}
	Consider the case where the number of heterogeneous workers selecting tasks at the NE satisfies $\sum_{m \in \mathcal{M}} N_m^{NE}(\boldsymbol{R}, \boldsymbol{Q}) = N$. If we choose the same $\boldsymbol{R}$ and $\boldsymbol{Q}$ in the BR case, then every high quality worker will participate by selecting one of the $M$ tasks in the BR case (i.e., $s_n \neq 0, \forall n \in \mc{N}_H$).
\end{mythm}

Theorem \ref{thmhigh} shows that even if the rewards are high enough to motivate all the workers, only high quality workers are guaranteed to participate in the BR case.
The detailed proof is given in Appendix \ref{app:thmhigh}.

Although we are not able to theoretically prove the convergence of CHE $N_m^{CHE}(\boldsymbol{R}, \boldsymbol{Q})$ to the NE $N_m^{NE}(\boldsymbol{R}, \boldsymbol{Q})$ under the heterogeneous workers case and the case of $\sum_{m \in \mathcal{M}} N_m^{NE}(\boldsymbol{R}, \boldsymbol{Q}) < N$, we have numerically validated this property through extensive simulations under various system parameters in Appendix \ref{app:results}. To sum up, the above theoretical and empirical studies show that the FR model is a limiting case of the BR model, when the average cognitive level $\tau$ approaches infinity.

\vspace{-3pt}
%==============================================================================
\section{Numerical Results} \label{sec:comparison}
%==============================================================================
\vspace{-3pt}

In this section, we present the numerical results to demonstrate the performance of our multimedia crowdsourcing solution. We first simulate the performance of Algorithm 1 in Section \ref{sec:alg}, and illustrate the requester's optimal rewards and quality requirements in Section \ref{sec:CHstg1}. Then, we further numerically compare the FR and BR models in Section \ref{sec:generalt}.

Unless specified otherwise, we will use the utility function $U_m(Q_m N_m) = u_m \ln (1 + {Q_m N_m})$ in the numerical results.

%\vspace{-3pt}
%%%%%%%%%%%%%%%%%%%%%%%%%%%%%%%%%%%%%%%%%%%%%%
\subsection{Performance Evaluation of Algorithm 1}
\label{sec:alg}

In this subsection, we present our simulation results of the low complexity Algorithm 1 to solve Problem 4. We show that Algorithm 1 can achieve a good performance, where the profit is close to that under optimal settings.

Here we randomly generate $M = 5, 10$, and $20$ tasks (small dimension scenarios), calculate the average profit under different $\alpha$ values, and compare with the optimal value (obtained by the exhaustive search) to show the near-optimality of our results.

\begin{figure}[t]
	\centering
	\includegraphics[height=9cm, trim = 0.5cm 0cm 0cm 0cm, clip = true]{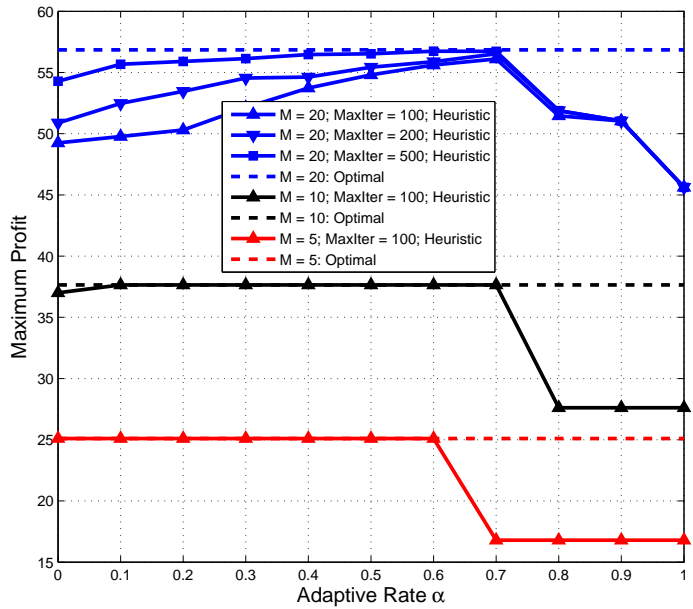}
	\caption{Maximum total profit with different $\alpha$ and $M$.} % and movement probability $= 0.4$
	\label{fig:grasp}
\end{figure}

When running Algorithm 1, we restrict $MaxIter = 100$ for $M = 5$ and 10 cases, while changing the value of $MaxIter$ for the case of $M = 20$. This is because $MaxIter = 100$ is not always enough to achieve convergence for $M = 20$ tasks. The value of $\alpha$ (the adaptive rate) also plays an important role. The results are summarized in Figure \ref{fig:grasp}, with detailed discussions as follows.

\begin{itemize}
	
	\item \emph{Small $\alpha$}: When $\alpha$ is less than 0.6, our proposed algorithm achieves the near-optimal performance. In particular, when $M$ is small (5 or 10), the solution computed by our proposed heuristic algorithm almost coincide with the optimal solution. When $M$ is large, our solution is close-to-optimal when we allow enough number of iterations (e.g., $MaxIter= 500$). 
	
	\item \emph{Large $\alpha$}: When $\alpha > 0.8$, our algorithm has a poor performance. This is because at each iteration, we restrict the feasible task candidate set to be the set that includes all new tasks with profit no smaller than $P^* = P_{l} + \alpha (P_{h} - P_{l})$.\footnote{Remind that the highest profit achieved by adding an additional task is denoted as $P_h$, and the lowest profit value is denoted as $P_l$ in Line 8 of Algorithm 1.} A large $\alpha$ value restricts the task candidate set, and the task choices cannot escape from a local optimal solution. Specifically, $\alpha = 1$ corresponds to a greedy algorithm which only adopts a new task which generates the highest profit at each iteration.

\end{itemize}

%The above analysis justifies our choice of $\alpha = 0.5$ and $MaxIter = 20M$ (where $M$ is the total number of tasks) in the paper, which achieves the best empirical trade off between the computation time and the solution performance.

Therefore, we choose $\alpha = 0.5$ and $MaxIter = 20M$ (where $M$ is the total number of tasks) to achieve the best empirical trade off between the computation time and the solution performance.

%\vspace{-3pt}
%%%%%%%%%%%%%%%%%%%%%%%%%%%%%%%%%%%%%%%%%%%%%%
\subsection{Optimal Rewards and Quality Requirements of the Requester} \label{sec:CHstg1}
In this subsection, we focus on the requester's optimal rewards and quality requirements in Stage I.

Note that the requester's optimization problem in (\ref{equorg}) is the same for both the FR and BR cases. Due to the progressive nature of Algorithm \ref{alg:CH} in determining the task selection, it is difficult to solve Problem \ref{prb:1} in closed-form in the BR case. Nevertheless, we can still numerically compute the Stage I optimal rewards $\boldsymbol{R}^{CHE*}$ and quality requirements $\boldsymbol{Q}^{CHE*}$. We denote the corresponding Stage II CHE as $\boldsymbol{N}^{CHE*} = \boldsymbol{N}^{CHE}(\boldsymbol{R}^{CHE*}, \boldsymbol{Q}^{CHE*})$ for simplicity.

To gain a deeper understanding of the result, we consider the case of $M = 3$ and choose the utility coefficients $\boldsymbol{u} = (30, 12, 8)$ and costs $\boldsymbol{c} = (2,1,3)$ for the sensing tasks. Then we compute $(\boldsymbol{R}^{CHE*},\boldsymbol{Q}^{CHE*})$ for workers with either homogeneous and heterogeneous quality capabilities and compare with the optimal task settings $(\boldsymbol{R}^{NE*},\boldsymbol{Q}^{NE*})$ in the FR case.

\subsubsection{Numerical Example on Homogeneous Worker Quality Capabilities} \label{sec:CH_homo}
In this case, the optimal quality requirement for any task $m$ is $Q_m^{CHE*} = q_L$. So we focus on characterizing how the optimal reward $R_m^{CHE*}$ changes in the total number of workers $N$ and the average cognitive level $\tau$.

\begin{figure*}[t]
	\hspace{-0.5cm}
	\centering
	\begin{minipage}[t]{0.45\linewidth}
		\includegraphics[width=8.5cm, trim = 0cm 0cm 0cm 0cm, clip = true]{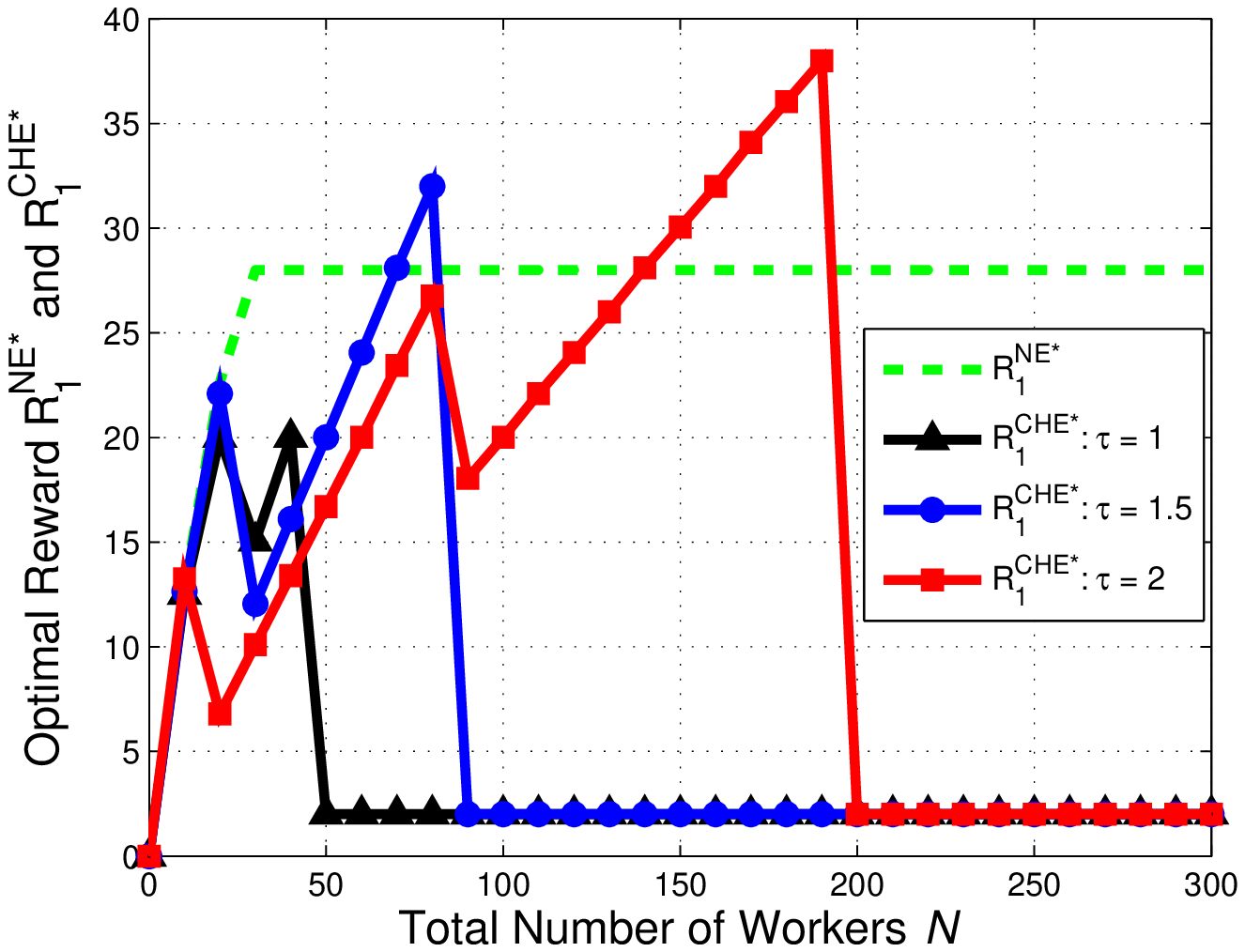}
		\vspace{-45pt}
		\caption{Homogeneous case: Optimal reward $R_1^{NE*}$ and $R_1^{CHE*}$ vs. $N$.} % for $I = 100$ and $c = 100$
		\label{fig:opt1}
	\end{minipage}
	\quad \quad 
	\begin{minipage}[t]{0.45\linewidth}
		\includegraphics[width=8.5cm, trim = 1.5cm 0cm 0cm 0cm, clip = true]{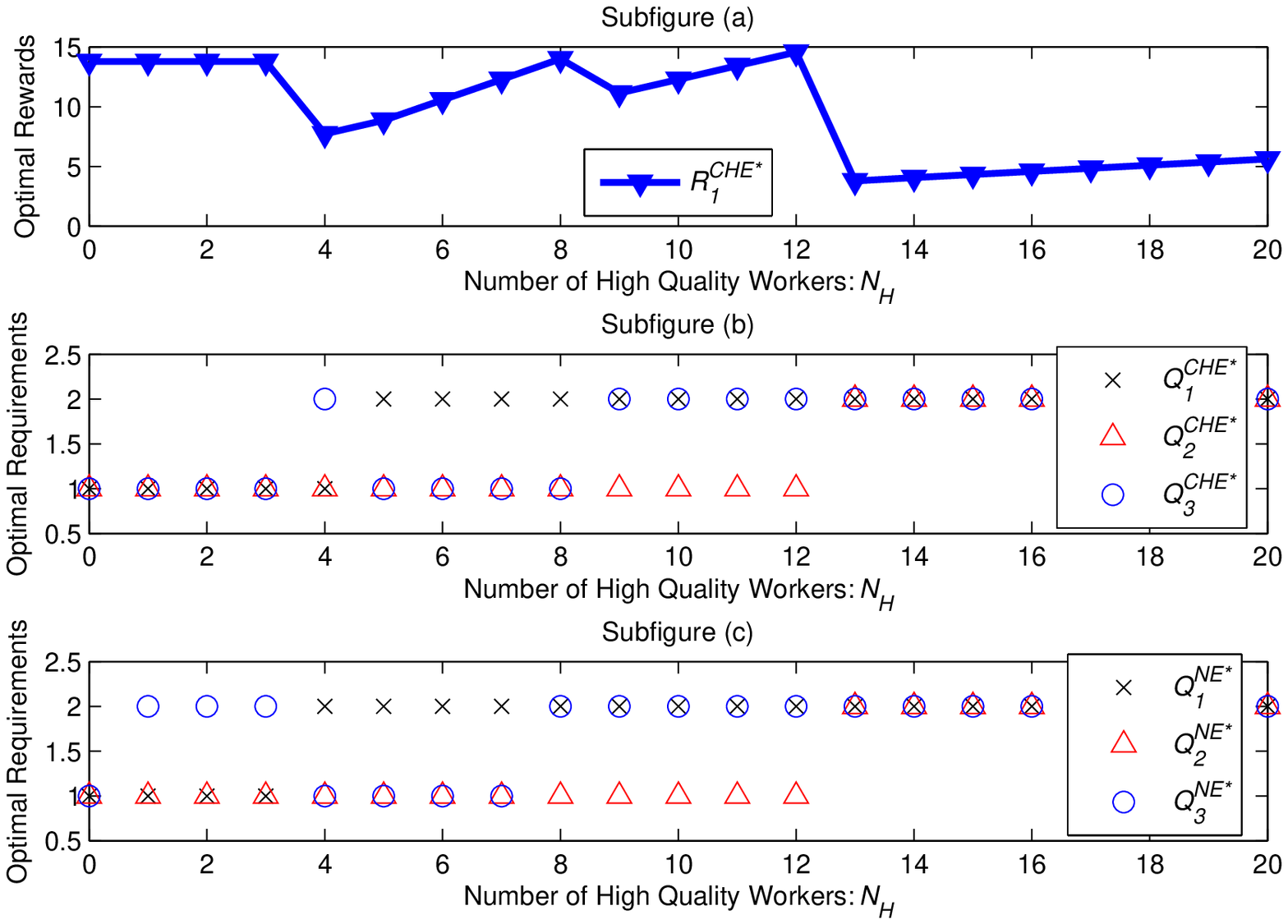}
		\vspace{-45pt}
		\caption{Heterogeneous case: $R_1^{CHE*}$, $\boldsymbol{Q}^{CHE*}$, and $\boldsymbol{Q}^{NE*}$ vs. $N_H$.} % for $I = 100$ and $c = 100$
		\label{fig:hetopt}
	\end{minipage}
	\vspace{-8pt}
\end{figure*}

In Fig. \ref{fig:opt1}, we first plot the optimal task 1 reward $R_1^{NE*}$ against $N$ in the FR model as a benchmark.\footnote{The optimal rewards for tasks 2 and 3 also follow similar patterns. We do not show them due to page limit.} As we can see, $R_1^{NE*}$ first increases with $N$ and then remains unchanged. With a small worker population (i.e., $N \leq 30$), the requester wants to attract more workers to complete the task. As $N$ further increases, however, the marginal revenue of attracting an additional worker becomes smaller and eventually cannot compensate the cost, and the requester will not increase the reward to attract them as $R_1^{NE*} = 28$.

Then we plot the optimal reward $R_1^{CHE*}$ against $N$ under different $\tau$ values.
As we can see, the optimal rewards do not monotonically increase in $N$. Consider the red curve of $\tau = 2$ as an example, and we can see different behaviors in several different ranges of N:
\begin{itemize}
	\item When $N$ increases from 20 to 80, $R_1^{CHE*}$ monotonically  increases as well. In this case, the requester wants to attract more workers to complete the task. 
	
	\item When $N$ reaches a threshold (i.e., $N=90$), the requester is able to reduce the reward, because she anticipates there are more low cognitive level workers who always participate as long as $R_m\geq c_m$. Then she can afford to provide a smaller reward just for the purpose of attracting the high level workers. 
	
	\item As $N$ becomes very large (e.g., $N \geq 200$), the optimal reward remains unchanged and equal to the cost $c_1$.
	In this case, as there are many \emph{level-0} workers in the population, it is enough for the requester just to attract and exploit these cheap labors to complete the task. 
\end{itemize}

Comparing the three curves with different $\tau$ values, a higher $\tau$ value (i.e., a more cognitive capable population) makes it more difficult for the requester to exploit the workers. For specifically,  the threshold for the requester to reduce the reward to $R_m^{CHE*} = c_m$ are $N =$ 50, 90, and 200 for $\tau =$ 1, 1.5, and 2, respectively.

\begin{myobs}
	The requester can take advantage of low cognitive level workers by setting smaller optimal rewards than the FR case, especially under a low average cognitive level and a large worker population.
\end{myobs}

\subsubsection{Numerical Example on Heterogeneous Worker Quality Capabilities} \label{sec:CH_het}
When the workers have heterogeneous quality capabilities, the impacts of $N$ and $\tau$ are similar to the homogeneous case. Hence, we will focus our study on how the number of high quality workers $N_H$ affects the requester's decisions. We fix $N = 20$, $\tau = 1.5$, $q_H = 2$, $q_L = 1$, and plot the requester's optimal decisions with respect to $N_H$ in Fig. \ref{fig:hetopt}.

In Fig. \ref{fig:hetopt}(a) and Fig. \ref{fig:hetopt}(b), we plot the optimal reward $R_1^{CHE*}$ and the optimal quality requirements $\boldsymbol{Q}^{CHE*}$ against $N_H$, respectively.\footnote{The analysis of $R_2^{CHE*}$ and $R_3^{CHE*}$ are similar. However, since the quality requirements for the three tasks change differently, the optimal rewards curves of three tasks mess up together. Here we only choose $R_1^{CHE*}$ to illustrate.} In Fig. \ref{fig:hetopt}(c), we plot the optimal quality requirements $\boldsymbol{Q}^{NE*}$ against $N_H$ to compare the FR and BR models.

We will first consider the CHE case in Fig. \ref{fig:hetopt}(a) and Fig. \ref{fig:hetopt}(b), depending on the value of $N_H$:
\begin{itemize}
	\item When $N_H$ is very small (i.e., $1 \leq N_H \leq 3$), Fig. \ref{fig:hetopt}(b) shows that it is optimal not to set any tasks with a high quality. 
	If the requester chooses a high quality requirement for a task, then the number of eligible workers is small. Instead, if she sets the quality requirement to be low, the number of eligible workers is much larger, which generates a higher profit than the high requirement case.
	
	\item When $N_H = 4$, the requester is able to take advantage of the high quality workers. However, $N_H$ is still not large enough, and the effect of having high number of workers also dominates the profit generation process.
	Fig. \ref{fig:hetopt}(a) shows that the requester chooses to set a high quality requirement for task 3 with the lowest utility parameter ($u_m$), while keeping the quality requirements of other two tasks low (so that they can attract enough workers). Fig. \ref{fig:hetopt}(a) shows that the reward for task 1 drops at $N_H=4$, as there are fewer workers that the request can incentivize.   
	
	\item As $N_H$ becomes larger (e.g., $N_H = 5$), the effect of high quality begins to dominate the profit generation process. It is better for the requester to choose the high quality requirement for tasks with high utility parameters. 
	Fig. \ref{fig:hetopt}(b) shows that as $N_H$ increases, eventually the requester sets the quality requirements to be high for all three tasks. Fig. \ref{fig:hetopt}(a) shows that the reward for task 1 drops a few times in this process, corresponding to the instances where the competitions among tasks increase and the number of workers potentially interested in task 1 reduces. 
	
\end{itemize}

When comparing Fig. \ref{fig:hetopt}(b) with Fig. \ref{fig:hetopt}(c), we note that the optimal task quality requirements under the BR case are similar to that under the FR case. This indicates that the requester can exploit workers' heterogeneity by setting different task quality requirements, regardless of the workers' reasoning capabilities (e.g., $4 \leq N_H \leq 12$ in Fig. \ref{fig:hetopt}(b) and Fig. \ref{fig:hetopt}(c)).\footnote{For example, Ctrip sets different photo-taking tasks for both professionals and amateurs.} 
When a single (i.e., high or low) quality capability workers dominate the platform (e.g., $13 \leq N_H \leq 20$ in Fig. \ref{fig:hetopt}(b) and Fig. \ref{fig:hetopt}(c)), however, the benefit of quality requirement differentiation disappears, because the number of the rest of the workers (with a different quality capability than the majority workers) is too small to make a difference.

When $N_H$ is very small (e.g., $1 \leq N_H \leq 3$ in Fig. \ref{fig:hetopt}(c)), consider the case where the requester chooses to set a high quality requirement for task 3. Under the FR model, all the high quality capability workers will focus on this task, and the profit is higher than the case where the requester set a low quality requirement for task 3.
Under the BR model, however, \emph{level-0} (including the high quality capability) workers always randomly select the tasks, hence setting a high quality requirement for task 3 will not improve the requester's profit.

\begin{myobs}
	Under both the FR and BR models, it is profitable for the requester to differentiate the tasks with different quality requirements, unless a single type of workers dominates the platform.
\end{myobs}

Now we focus on how the quality capabilities $q_H$ and $q_L$ impact the requester and workers. We fix $N = 40$, $N_H = 20$, $\tau = 1.5$, and $q_H = 10$, and change $q_L$ from 1 to 10 (i.e., $q_L \in [1, 10)$). We have the results as follows.
\begin{itemize}
	%\item Quality requirement: For both the FR and BR models, the optimal quality requirement for low requirement tasks equals $q_L$, hence increases with the value of $q_L$, while the high requirement will remain unchanged as $q_H$ (from Proposition 2 and exhaustive search).
	
	\item Workers' payoff: In the FR model, from Proposition 2, the reward just compensates every participating worker's cost, hence the payoff for every worker remains unchanged as zero. In the BR model, the optimal reward still cannot compensate low cognitive level workers' cost, where each worker will encounter a negative payoff.
	
	\item Requester's profit: For both the FR and BR models, as $q_L$ increases, the requester's profit will also increase. This is because these low quality capability workers are now increasing their capabilities, hence the requester can make more revenue from these workers.
\end{itemize}

\vspace{-3pt}
%%%%%%%%%%%%%%%%%%%%%%%%%%%%%%%%%%%%%%%%%%%%%%
\subsection{Profit Comparison of the FB and BR Models} \label{sec:generalt}

Now we focus on the profit comparison between the FR and BR models. We will show that the requester can take advantage of the BR model and obtain a higher level of profit.

We choose the values of $u_m$ and $c_m$ same as that in Section \ref{sec:CHstg1} with $M = 3$ tasks. In the FR model, we compute the near-optimal rewards $\boldsymbol{R}^{N-OPT}$, quality requirements $\boldsymbol{Q}^{N-OPT}$ (based on Algorithm \ref{alg:m1}), and the corresponding NE $N_m^{N-OPT} = N_m^{NE}(\boldsymbol{R}^{N-OPT}, \boldsymbol{Q}^{N-OPT})$ for each $m$, or $(N_m^{N-OPT}, \forall m \in \mc{M})$ for simplicity. In the BR model, we set $\tau = 5$. In the following, we will consider two different cases, depending on whether $\bs{R}$ and $\bs{Q}$ are the same in both models.

\begin{figure*}[t]
	\hspace{-0.5cm}
	\centering
	\begin{minipage}[t]{0.45\linewidth}
		\includegraphics[width=8.5cm, trim = 0cm 0cm 0cm 0cm, clip = true]{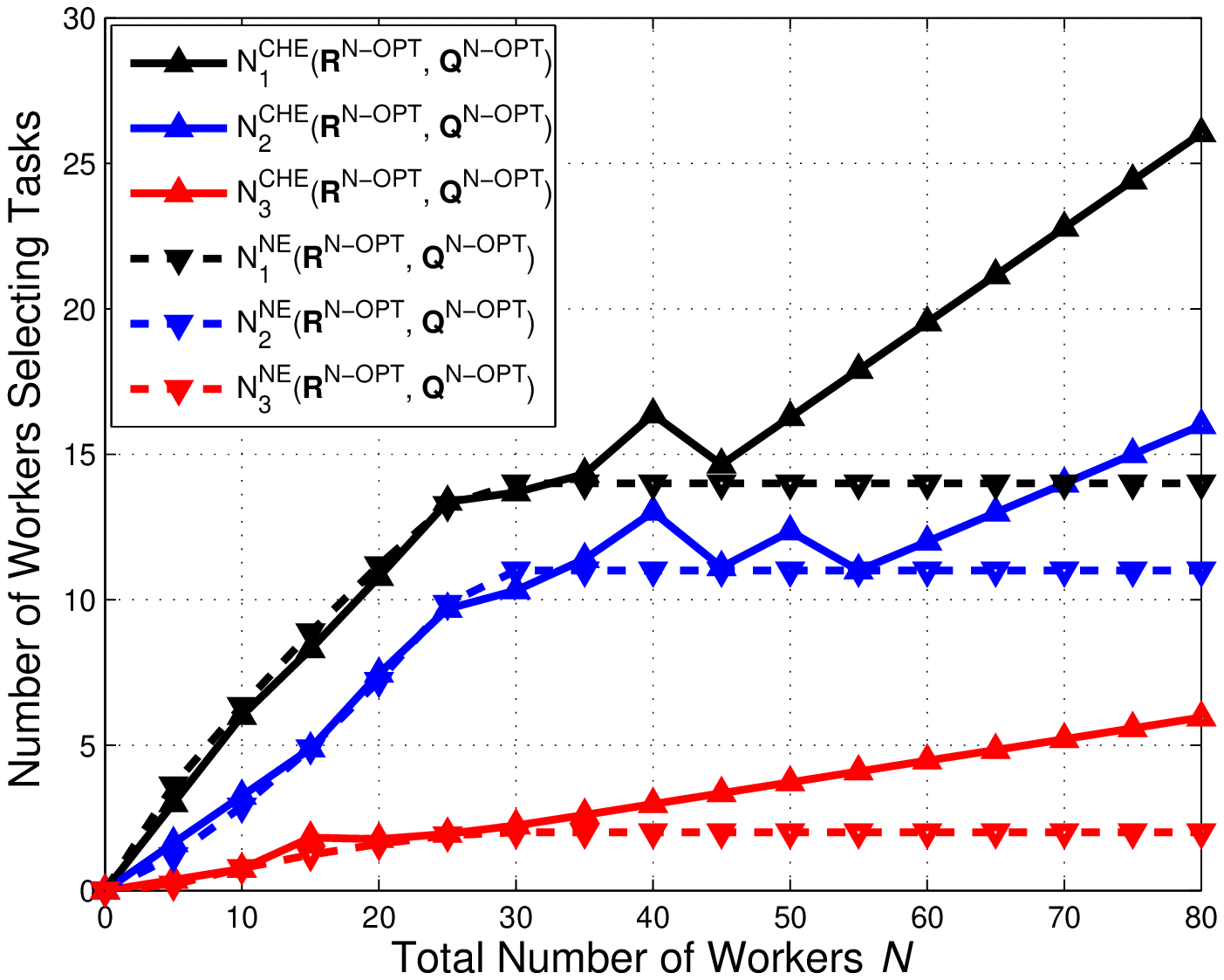}
		\vspace{-45pt}
		\caption{Selecting number in FR and BR vs. $N$.} % for $I = 100$ and $c = 100$
		\label{fig:NE1}
	\end{minipage}
	\quad \quad 
	\begin{minipage}[t]{0.45\linewidth}
		\includegraphics[width=8.5cm, trim = 0cm 0cm 0cm 0cm, clip = true]{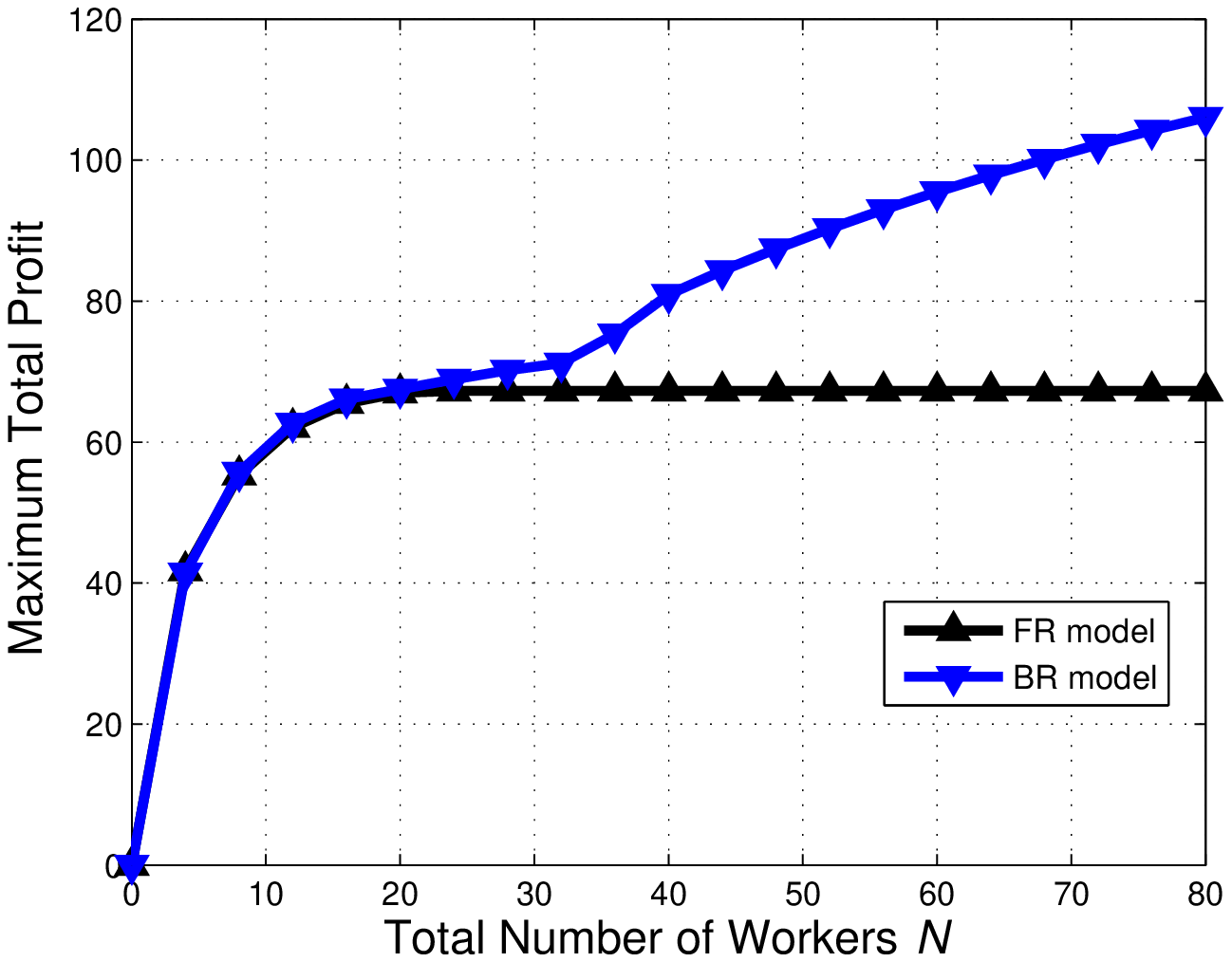}
		\vspace{-45pt}
		\caption{Maximum profit in FR and BR vs. $N$.} 
		\label{fig:NE3}
	\end{minipage}
\vspace{-5pt}
\end{figure*}

\subsubsection{The requester sets the same rewards and quality requirements}
In the first scenario, we assume that the requester sets the same rewards $\boldsymbol{R}^{N-OPT}$ and quality requirements $\boldsymbol{Q}^{N-OPT}$ in Stage I (both for the FR and BR models), and plot the number of workers selecting tasks for both models in Fig. \ref{fig:NE1}.
This allows us to achieve a fair comparison in terms of the workers' behaviors in Stage II. 

When $N$ is small (i.e., $N \leq 25$), Fig. \ref{fig:NE1} shows that the number of workers $N_m$ for each task $m$ is similar under NE and CHE (e.g., the two black curves of $N_1^{N-OPT}$ and $N_1^{CHE}(\boldsymbol{R}^{N-OPT}, \boldsymbol{Q}^{N-OPT})$), and increases with $N$. 
Under a small worker population, few workers belonging to the same level due to the assumption of Poisson distribution in Section \ref{sec:CH}, hence different workers make their decisions sequentially as in Algorithm \ref{alg:CH}. This means that higher level workers will be able to adjust their choices by responding to the lower-level workers' choices. That is to say, if the number of workers selecting task $m$ exceeds $N_m^{N-OPT}$, higher level workers will no longer select the same task. As a result, given the same rewards in the FR and BR models, the $N_m^{N-OPT}$ and $N_m^{CHE}(\boldsymbol{R}^{N-OPT}, \boldsymbol{Q}^{N-OPT})$ in Stage II will be similar.

As $N$ becomes large enough (i.e., $N \geq 55$), Fig. \ref{fig:NE1} shows that the $N_m^{N-OPT}$ in the FR model no longer changes (as the task has attracted enough workers when the marginal revenue equals the cost). Meanwhile, $N_m^{CHE}(\boldsymbol{R}^{N-OPT}, \boldsymbol{Q}^{N-OPT})$ in the BR model continues to increase and the value is larger than $ N_m^{N-OPT}$. The reason is that before a particular worker selects the task $m$, the number of workers selecting the same task may be close but still less than $N_m^{N-OPT}$, and that worker believes he will gain a positive payoff if he is the only new worker to select this task. However, there can be multiple workers at the same level going through the same reasoning process, and multiple workers may end up selecting the same task. As a result,  $N_m^{CHE}(\boldsymbol{R}^{N-OPT}, \boldsymbol{Q}^{N-OPT})$ will exceed $N_m^{N-OPT}$.
When $N$ increases, every cognitive level contains more workers, so the difference between $N_m^{CHE}(\boldsymbol{R}^{N-OPT}, \boldsymbol{Q}^{N-OPT})$ and $N_m^{N-OPT}$ becomes larger.

\begin{myobs}
	Given the same task settings ($\bs{R}$ and $\bs{Q}$), the number of workers selecting tasks in the BR model increases unboundedly with the total number of workers, while it will saturate in the FR model.
\end{myobs}

\subsubsection{The requester can set different profit maximizing rewards and quality requirements}
Next we consider a second scenario where the requester can freely choose her profit maximizing rewards and requirements in Stage I to maximize her payoff, which leads to the corresponding NE and CHE. Fig. \ref{fig:NE3} illustrates the maximum total profits in both the FR and BR models. 

When $N$ is small (i.e., $N \leq 20$), the maximum profits in both models are the same and increase with $N$.
The explanation is similar to that in Fig. \ref{fig:NE1} for the small $N$ case.

As $N$ becomes large (i.e., $N \geq 25$), Fig. \ref{fig:NE3} shows that the profit in the FR model will not further increase because the near-optimal reward $\boldsymbol{R}^{N-OPT}$, quality requirement $\boldsymbol{Q}^{N-OPT}$ and the NE $\boldsymbol{N}^{N-OPT}$ do not further increase in N. However, the profit in the BR model goes up unboundedly with the total number $N$.
This is because as $N$ increases, there are more \emph{level-0} workers, hence eventually the requester can simply set $R_m=c_m$ and relies on the increasing number of \emph{level-0} workers to accomplish these tasks. 
\begin{myobs}
	The profit in the BR model increases unboundedly with the number of workers, while it saturates in the FR model.
\end{myobs}

%\section{Conclusion}
\vspace{-2pt}
%==============================================================================
\section{Conclusions and Future Work} \label{sec:concl}
\vspace{-2pt}
%==============================================================================
In this paper, we presented the first study regarding how the bounded cognitive rationality (i.e., the cognitive limits of reasoning) affects the incentive mechanism design in multimedia crowdsourcing systems. 
In the fully rational benchmark case, we derived the requester's optimal rewards and quality requirements, as well as the workers' Nash equilibrium selections. Due to the high computational complexity in the heterogeneous workers' case, we proposed a low-complexity heuristic algorithm to compute the near-optimal solution.
In the bounded rational case, we modeled the workers' belief formation and progressive decision process based on the cognitive hierarchy theory in behavioral economics. We analyzed how the requester can improve her profit by taking advantage of the workers' imperfect reasoning.
Simulation results showed that the requester is more likely to obtain a higher profit under a larger worker population with a lower average cognitive level.
%In this paper, we assumed that the workers have the same task costs. 

For the future work, we will consider the impact of bounded cognitive rationality in more practical system scenarios. For example, it is interesting to consider rewards that depend on both the workers' sensing efforts and quality capabilities, and worker costs that are both task and worker dependent. %Then we will find how the workers' bounded cognitive rationality influences workers' payoffs and requester's profit.
Also, we will conduct field trials to validate our theory, e.g., estimating the value of average cognitive level $\tau$ for different types of crowdsourcing applications.

%%%%%%%%%%%%%%%%%%%%%%%%%%%%%%%%%%%%%%%%%%%%%%%%%%%%%%%%%%%%%%%%%
%\bibliographystyle{unsrt}
\bibliographystyle{IEEEtran}
\bibliography{IEEEabrv,mybibfile}

%============================================================================
%\newpage
%\setcounter{page}{1}
\textwidth 6.65in
\renewcommand{\baselinestretch}{1.4}
%%%%%%%%%%%%%%%%%%%%%%%%%%%%%%%%%%%%%%%%%%%%%%%%%%%%%%%%%%%%%%%%%%
%\begin{center}
%	\Large{\textbf{Congestion-Aware Distributed Network Selection \\ for Cellular and Wi-Fi Integration}} \\
%	\Large{\textbf{Multimedia Crowdsourcing with Bounded Rationality: \\ A Cognitive Hierarchy Perspective}} \\
%	\small{Qi Shao, Man Hon Cheung, and Jianwei Huang}
%\end{center}
\normalsize
\appendix
%
%============================================================================
\subsection{Proof of Proposition \ref{prphet}} \label{app:prphet}
We need to prove that constraints (\ref{eq:prphet}a)-(\ref{eq:prphet}f) are sufficient and necessary conditions in two directions. 
\begin{itemize}
	\item [I)] We first prove the ``if'' part:
	
	\begin{itemize}
		\item [a)] First, we discuss the task selections of workers with high quality capabilities. We define a task set $\mc{M}_1 = \{m \in \mc{M}: N_{H,m}^{NE} > 0\}$ containing tasks which are selected by high quality workers. Then we define $\mc{M}_2 = \{m \in \mc{M}: N_{H,m}^{NE} = 0\} = \mc{M} \backslash \mc{M}_1$ containing tasks which are not selected by any high quality workers.
		
		For a high quality capability worker, if he selects a task $m \in \mc{M}_1$, from constraint (\ref{eq:prphet}a) he will gain a payoff of $\frac{R_m}{N_{H,m}^{NE} + N_{L,m}^{NE}} - c_m = \lambda_1 \geq 0$. 
		
		If he changes to select a task $m \in \mc{M}_2$, from constraint (\ref{eq:prphet}a) he will gain a payoff of $\frac{R_m}{N_{H,m}^{NE} + N_{L,m}^{NE}} - c_m \leq \lambda_1$.
		
		If not all the high quality workers participate (i.e., $\sum_{m \in \mathcal{M}} N_{H,m}^{NE} < N_H$), then we have $\lambda_1 = 0$ from constraint (\ref{eq:prphet}b), and a high quality capability worker has no incentive to select a task.
		
		Hence, given constraints constraints (\ref{eq:prphet}a)-(\ref{eq:prphet}f), a worker with high quality capability has no incentive to change his task selection strategy.

		\item [b)] Then, we discuss the task selections of workers with low quality capabilities.
		
		For a low quality capability worker, he has no incentive to select a task $m \in \mc{M}_H$, because he will always receive a payoff of $-c_m < 0$.
		
		The analysis for constraints (\ref{eq:prphet}d) and (\ref{eq:prphet}e) are similar to (\ref{eq:prphet}a) and (\ref{eq:prphet}b).
		
		Hence, given constraints constraints (\ref{eq:prphet}c)-(\ref{eq:prphet}e), a worker with low quality capability has no incentive to change his task selection strategy.

	\end{itemize}
	Overall, if constraints (\ref{eq:prphet}a)-(\ref{eq:prphet}f) are all satisfied, $(\bs{N}^{NE}_H, \bs{N}^{NE}_L)$ is an NE.

	\item [II)] We then proof the ``only if'' part:
	
	If $(\bs{N}^{NE}_H, \bs{N}^{NE}_L)$ is an NE, no workers would like to deviate from the current state.
	\begin{itemize}
		\item [a)]	For a high quality worker, if he selects task $m \in \mc{M}_1$ and has no incentive to change his strategy, because the payoff of selecting task $m$ is no lower than that of choosing other strategies. Hence, we have $\lambda_1 = \frac{R_m}{N_{H,m}^{NE} + N_{L,m}^{NE}} - c_m \geq 0, \fa{m}{\mc{M}}_1$. Then we have constraint (\ref{eq:prphet}a) because $N_{H,m}^{NE} \geq 0$. If not all high quality capabilities workers participate, then we have constraint (\ref{eq:prphet}b) so that they will not change their strategies. If all the high quality capabilities workers participate, then constraint (\ref{eq:prphet}b) is always satisfied.
		
		\item [b)] 	For a low quality capability worker, he has no incentive to select a task $m \in \mc{M}_H$, hence we have constraint (\ref{eq:prphet}c).
		
		The analysis for constraints (\ref{eq:prphet}d) and (\ref{eq:prphet}e) are similar to (\ref{eq:prphet}a) and (\ref{eq:prphet}b). Constraint (\ref{eq:prphet}f) is the model setting of the paper.
	\end{itemize}

	Overall, under the NE $(\bs{N}^{NE}_H, \bs{N}^{NE}_L)$, constraints (\ref{eq:prphet}a)-(\ref{eq:prphet}f) are all satisfied.
\end{itemize}

%\hfill \QEDclosed

\subsection{Proof of Proposition \ref{prphetopt}} \label{app:prphetopt}

We need to prove that (\ref{eq:hetopt}a)-(\ref{eq:hetopt}c) are necessary conditions for Problem (Profit-Bilevel). In other words, if (\ref{eq:hetopt}a)-(\ref{eq:hetopt}c) are not satisfied, constraints in Problem (Profit-Bilevel) will be violated, or the optimal value will not be achieved.

Firstly, given the rewards $\bs{R}$ and the task sets $\mc{M}_H$ and $\mc{M}_L$, the number of workers selecting tasks under the NE $\bs{N}_H^{NE}$ and $\bs{N}_L^{NE}$ are determined. Then we have constraint (\ref{eq:hetopt}a) to determine the optimal quality requirements to maximize the requester's profit.

Given constraint (\ref{eq:hetopt}b), we have $\lambda_1 = \lambda_2 = 0$ in constraints (\ref{eq:prphet}a)-(\ref{eq:prphet}f). And we will focus on proving $\lambda_1 = 0$ case, because the proof for $\lambda_2 = 0$ is similar. There are two cases that we need to consider: $\lambda_1 < 0$ and $\lambda_1 > 0$.
\begin{itemize}
	\item If $\lambda_1 < 0$, then constraint (\ref{eq:prphet}f) is violated.
	\item If $\lambda_1 > 0$, we can always find a $\hat{\lambda_1} > 0$ satisfying $\lambda_1 > \hat{\lambda_1} \geq 0$, and we have 
	\vspace{-2pt}
	\begin{displaymath}
	\hat{R_m} = \left(c_m + \hat{\lambda_1}\right) \left(N_{L,m}^{NE} + N_{H,m}^{NE}\right) < R_m = \left(c_m + \lambda_1\right) \left(N_{L,m}^{NE} + N_{H,m}^{NE}\right), \forall m \in \mc{M}.
	\end{displaymath}
	Then we have 
	\begin{displaymath}
	U_m (Q_m (N_{L,m}^{NE} + N_{H,m}^{NE})) - \hat{R_m} > U_m (Q_m (N_{L,m}^{NE} + N_{H,m}^{NE})) - R_m, \forall m \in \mc{M}.
	\end{displaymath}
	which implies that $\lambda_1 > 0$ is not the optimal solution to the problem. 
\end{itemize}

Thus, constraint (\ref{eq:hetopt}b) is also necessary for Problem (Profit-Bilevel). And we have constraint (\ref{eq:hetopt}c) because both the number of high quality capabilities workers and the total number of workers are non-negative and limited.

%\hfill \QEDclosed

\subsection{Proof of Theorem \ref{thmall}} \label{app:thmall}
Firstly, we prove that under the case when workers are with homogeneous quality capabilities, if the workers all participate in the FR  model, they will all participate in the BR model. Then we prove that the CHE converges to the NE as $\tau \to \infty$.

Given the rewards $\bs{R}$ and the quality requirements $\bs{Q}$, we have $\sum_{m \in \mathcal{M}} N_m^{NE}(\boldsymbol{R}, \boldsymbol{Q}) = N$, where all the $N$ workers are with homogeneous quality capabilities. From Proposition \ref{prphet}, for a task $m \in \mc{M}$ with $R_m > 0$, there will be $N_m = \frac{R_m}{c_m + \lambda}$ workers selecting task $m$, where $\lambda \geq 0$ is every worker's payoff. And because of the positive rewards, in the BR model, all the \emph{level-0} workers will participate.

Then we prove that for a \emph{level-$k$} $(k\geq 1)$ worker, if he believes that all lower level workers participate, he will also select a task and participate.

For a \emph{level-$k$} worker, there are $\widetilde{N_m}(k)$ workers selecting task $m$ in his belief and $\sum_{m \in \mathcal{M}} \widetilde{N_m}(k) = N$. Then there must be at least a task $m \in \mc{M}$ such that $\frac{R_m}{\widetilde{N_m(k)}} - c_m \geq \lambda \geq 0$ and this worker will select task $m$. If not, then we have $\widetilde{N_m}(k) > \frac{R_m}{c_m + \lambda}, \fa{m}{\mc{M}}$, which leads to a contradiction that $\sum_{m \in \mathcal{M}} \widetilde{N_m}(k) > \sum_{m \in \mathcal{M}} \frac{R_m}{c_m + \lambda} = N$. Hence, all the \emph{level-$k$} workers are willing to select a task and participate in the BR case. And this completes the proof that all the workers will participate in the BR model.

Then we define $n_m(k) = N f(k) e(k,m)$ as the number of \emph{level-$k$} workers joining task $m$, and $N_m(k) = \sum_{j=0}^{k} n_m(j)$ as the total number of workers from \emph{level-0} to \emph{level-$k$} selecting task $m$. 
Thus, we have $N_m^{CHE}(\bs{R}, \bs{Q}) = N_m(\infty)$ and $n_m(k) = N_m(k) - N_m(k-1)$, $k \geq 1$. 

If $N_m(k)$ reaches or exceeds $N_m^{NE}(\bs{R}, \bs{Q})$, then workers with higher level than $k$ will never select task $m$. Because for \emph{level-$(k+i)$} ($i \geq 1$) workers, we have
\vspace{-2pt}
\begin{displaymath}
\begin{aligned}
\frac{N_m({{k}+i-1})}{\sum_{j=0}^{{k} + i - 1} f(j)} \geq \frac{N_m(k)}{\sum_{j=0}^{k + i - 1} f(j)} > N_m(k) \geq N_m^{NE} (\bs{R}, \bs{Q}),\\
E(k+i,m) = \frac{R_m}{\frac{N_{{k} +i -1}}{\sum_{j=0}^{{k} +i -1} f(j)}} - c_m  < \frac{R_m}{N_m^{NE} (\bs{R}, \bs{Q})} - c_m < \lambda.
\end{aligned}
\vspace{-2pt}
\end{displaymath}

In other words, all the workers with levels higher than $k$ will not select task $m$, because there exists at least one task with payoff higher than $\lambda$ and we have $N_m^{CHE}(\bs{R}, \bs{Q}) = N_m(k)$.

Before proving the convergence property, we need to have the following lemmas for preparation.

\begin{mylem}
	\label{lemt}
	In the BR model, level-$\lfloor \tau \rfloor$ accounts for the most fraction with $\frac{\tau ^ {\lfloor \tau \rfloor}}{\lfloor \tau \rfloor !} e^{-\tau}$. 
\end{mylem}
\begin{IEEEproof}
	Since $f(k) = \frac{\tau ^ k }{k!} e^{-\tau}$, we can have $\frac{f(k)}{f(k-1)} = \frac{\tau}{k}$, in other words, for $k \leq \lfloor \tau \rfloor $, we always have $\frac{f(k)}{f(k-1)} = \frac{\tau}{k} \geq 1$; for $k \geq \lfloor \tau + 1 \rfloor $, we always have $\frac{f(k)}{f(k-1)} = \frac{\tau}{k} < 1$. Thus, if $\tau$ is an integer, level $\tau$ and $(\tau -1)$ account for the most fraction, with $f(\tau) = f(\tau - 1) = \frac{\tau ^ {\tau}}{\tau !} e^{-\tau}$; if $\tau$ is not an integer, level-$\lfloor \tau \rfloor$ accounts for the most fraction with $\frac{\tau ^ {\lfloor \tau \rfloor}}{\lfloor \tau \rfloor !} e^{-\tau}$.
\end{IEEEproof}

%%%%%%%%%%%%%lemma2
\begin{mylem}
	\label{lem0}
	$\lim_{\tau \to \infty} \frac{\tau ^ {\lfloor \tau \rfloor}}{\lfloor \tau \rfloor !} e^{-\tau} = 0$.
\end{mylem}
\begin{IEEEproof}
	Define non-negative integer $n = \lfloor \tau \rfloor$ and sequence \{$x_n = \frac{n^n}{n!} e^{-n}$\}. Then we have
	\vspace{0pt}
	\begin{displaymath}
	\begin{aligned}
	\frac{x_{n+1}}{x_{n}} &= \frac{e^{-(n+1)} \cdot (n+1) ^ {(n+1)} \cdot n! }{e^{-n} \cdot n ^ n \cdot (n+1)! }\\
	&=  \frac{1}{e} \cdot (\frac{n+1}{n}) ^ {n}.
	\end{aligned}
	\end{displaymath}
	
	Since $x_n > 0$ and $(1+\frac{1}{n})^n < e$, we have $\frac{x_{n+1}}{x_{n}} < 1$. Hence, the sequence \{$x_n$\} is monotonically decreasing with a lower bound 0. Thus, the sequence has a limit. For large $n$ value, we use Stirling's approximation (i.e., $n!  \sim \sqrt{2 \pi n} (\frac{n}{e})^n$) and
	\vspace{-3pt}
	\begin{displaymath}
	\begin{aligned}
	\lim_{n \to \infty} x_n &= \lim_{n \to \infty} \frac{n^n}{n!} e^{-n} \\
	&= \lim_{n \to \infty} \frac{n^n}{\sqrt{2 \pi n} (\frac{n}{e})^n} e^{-n}\\
	&= 0.
	\end{aligned}
	\vspace{-3pt}
	\end{displaymath}
	
	Since $n = \lfloor \tau \rfloor$, we have $\lim_{\tau \to \infty} \frac{\tau ^ {\lfloor \tau \rfloor}}{\lfloor \tau \rfloor !} e^{-\tau} = \lim_{\tau \to \infty} \frac{\lfloor \tau \rfloor ^ {\lfloor \tau \rfloor}}{\lfloor \tau \rfloor !} e^{-\lfloor \tau \rfloor} = \lim_{n \to \infty} x_n = 0$.
\end{IEEEproof}

From Lemma \ref{lemt}, we have $N_m^{CHE}(\bs{R}, \bs{Q}) - N_m^{NE}(\bs{R}, \bs{Q}) < n_m(k) \leq \frac{\tau ^ {\lfloor \tau \rfloor}}{\lfloor \tau \rfloor !} e^{-\tau} N$.

Then we calculate $N_m^{CHE}(\bs{R}, \bs{Q}) - N_m^{NE}(\bs{R}, \bs{Q})$, and suppose for task $m \in \mathcal{M}_1$, we have $N_m^{CHE}(\bs{R}, \bs{Q}) - N_m^{NE}(\bs{R}, \bs{Q}) > 0$, and  $N_m^{CHE}(\bs{R}, \bs{Q}) - N_m^{NE}(\bs{R}, \bs{Q}) < 0$ for task $m \in \mathcal{M}_2$. Note that
\begin{displaymath}
\begin{aligned}
|\mathcal{M}_1| + |\mathcal{M}_2| & \leq M,\\
\sum_{m \in \mathcal{M}_1} |N_m^{CHE}(\bs{R}, \bs{Q}) - N_m^{NE}(\bs{R}, \bs{Q})| & = \sum_{m \in \mathcal{M}_2} |N_m^{CHE}(\bs{R}, \bs{Q}) - N_m^{NE}(\bs{R}, \bs{Q})|.
\end{aligned}
\end{displaymath}

Thus, we have
\begin{displaymath}
\underset{m \in \mathcal{M}_2}{\max} |N_m^{CHE}(\bs{R}, \bs{Q}) - N_m^{NE}(\bs{R}, \bs{Q})| \leq \sum_{m \in \mathcal{M}_1} |N_m^{CHE}(\bs{R}, \bs{Q}) - N_m^{NE}(\bs{R}, \bs{Q})| \leq  \frac{\tau ^ {\lfloor \tau \rfloor}}{\lfloor \tau \rfloor !} e^{-\tau} M_1 N,
\end{displaymath}
and from Lemma \ref{lem0}, we have 
\begin{displaymath}
\lim_{\tau \to \infty} |N_m^{CHE}(\bs{R}, \bs{Q}) - N_m^{NE}(\bs{R}, \bs{Q})| = 0, \forall m \in \mathcal{M}.
\end{displaymath} 
In other words, the CHE converges to NE as $\tau$ approaches $\infty$.

%\hfill \QEDclosed

\subsection{Proof of Theorem \ref{thmhigh}} \label{app:thmhigh}
There are two cases of heterogeneous workers selecting tasks: 
\begin{itemize}
	\item All the high quality capability workers focus on high quality requirement tasks in the FR model.
	\item The high quality capability workers also focus on low quality requirement tasks in the FR model.
\end{itemize}

In the first case, given the same rewards and quality requirements in the BR model, \emph{level-0} high quality capability workers will always participate and randomly select the tasks (including low quality requirement tasks). If they all focus on the high quality requirement tasks, all these workers will get a payoff of $\lambda_1$ from Proposition \ref{prphet}. Similar to the proof of Theorem \ref{thmall}, for a \emph{level-$k$} worker, there must be at least one task $m \in \mc{M}_1$ with payoff $\frac{R_m}{\widetilde{N_m}(k)} - c_m \geq \lambda_1 \geq 0$. If not, we have $\sum_{m \in \mathcal{M}_1} \widetilde{N_m}(k) > N_H$ which is a contradiction. Hence all the high quality workers will participate. 

In the second case, from Proposition \ref{prphet}, we have $\lambda_1 = \lambda_2 \geq 0$. And a worker (regardless of the quality capability) will receive a payoff of $\lambda = \lambda_1 = \lambda_2$ as he selects task $m \in \mc{M}$. Similarly, for a \emph{level-$k$} worker with high quality capability, there must be at least one task $m \in \mc{M}$ with payoff $\frac{R_m}{\widetilde{N_m}(k)} - c_m \geq \lambda \geq 0$. If not, we have $\sum_{m \in \mathcal{M}} \widetilde{N_m}(k) > N$ which is a contradiction. Hence all the high quality workers will participate. 

For low level quality capability workers, they can only select low requirement tasks, and there is no guarantee they will always participate in the BR case.
%\hfill \QEDclosed

%
%
%
\begin{figure}[t]
	\centering
	\includegraphics[height=7cm, trim = 0.5cm 0cm 0cm 0cm, clip = true]{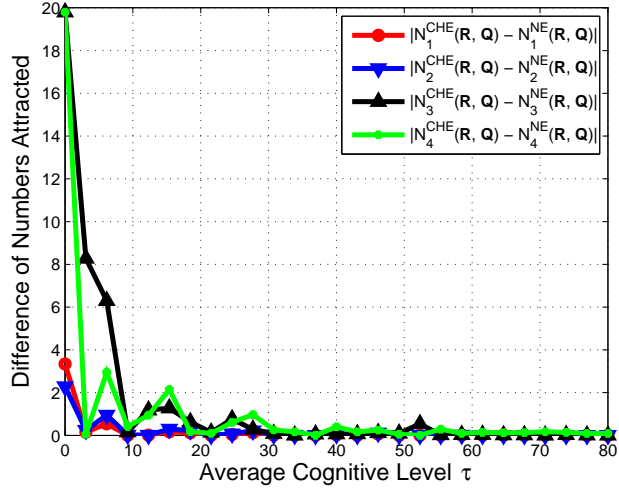}
	\vspace{-20pt}
	\caption{Difference of two models vs. $\tau$.} % for $I = 100$ and $c = 100$
	\label{fig:3}
	\vspace{-12pt}
\end{figure}

\subsection{Numerical Results on Convergence for $\sum_{m \in \mathcal{M}} N_m^{NE}(\bs{R}, \bs{Q}) < N$ Case and Heterogeneous Case} \label{app:results}
The convergence properties can be shown by the difference of $|N_m^{CHE}(\bs{R}, \bs{Q}) - N_m^{NE}(\bs{R}, \bs{Q})|$ when setting the same rewards $\bs{R}$ and quality requirements $\bs{Q}$ in the FR and BR models, and Fig. \ref{fig:3} is a simple example when setting $\bs{c} = (1,2,1.5,2)$, $\bs{R} = (15,20,20,30)$, $\bs{Q} = (1,q_H,q_H,1)$, $N=100$, and $N_H = 30$.

For different system parameters, we fix costs as $\bs{c} = (1,2,1.5,2)$, change the rewards $\bs{R}$, quality requirements $\bs{Q}$, the number of high quality capability workers $N_H$ and the total number $N$, and compare the difference between CHE and NE (i.e., $\underset{m \in \mc{M}}{\max} \{|N_m^{CHE}(\bs{R}, \bs{Q}) - N_m^{NE}(\bs{R}, \bs{Q})|\}$) under different $\tau$ values.
Table \ref{table:1} and \ref{table:2} list some of the results, and the difference between CHE and NE decreases with $\tau$.

%%%%%%%%%%%%%%%%%%%%%%% ttt %%%%%%%%%%%%%%%%%%%%%%%%%%%%%%%%%%%
\begin{table} [t]
	%\renewcommand{\arraystretch}{1.3}
			\small
	\caption{$N = 50$, $N_H = 15$: ${\max} \{|N_m^{CHE}(\bs{R}, \bs{Q}) - N_m^{NE}(\bs{R}, \bs{Q})|\}$} % (a)-(d)
	\label{table:1}
	\vspace{-5pt}
	\centering
	\ra{1.25}
	\begin{tabular}{@{}cccccc@{}} 
		%\begin{tabular}{|c|c|c|}
		\hline
		\multirow{2}{4cm}{\textbf{Rewards \& Quality requirements}} & \multicolumn{5}{c}{\textbf{$\tau$ Value}} \\
		\cline{2-6}
		& 5 & 10 & 20 & 40 & 80\\
		
		\hline
		
		$\bs{R} = (5, 20, 15, 10)$, $\bs{Q} = (q_H, 1, 1, 1)$ & 3.8772 & 2.2020 & 1.0131 & 0.2396 & 0.0495\\
		
		\hline

		$\bs{R} = (12, 18, 15, 8)$, $\bs{Q} = (q_H, 1, 1, q_H)$ & 3.0778 & 2.0937 & 1.6583 & 1.0584 & 0.6854\\
		
		\hline

		$\bs{R} = (20, 14, 17, 6)$, $\bs{Q} = (1, 1, q_H, 1)$ & 4.1226 & 2.7792 & 1.0168 & 0.6666 & 0.3158\\

		\hline
		
	\end{tabular}
	\vspace{-5pt}
\end{table}

\begin{table} [t]
	\small
	%\renewcommand{\arraystretch}{1.3}
	\caption{$N = 200$, $N_H = 40$: ${\max} \{|N_m^{CHE}(\bs{R}, \bs{Q}) - N_m^{NE}(\bs{R}, \bs{Q})|\}$} % (a)-(d)
	\label{table:2}
	\vspace{-5pt}
	\centering
	\ra{1.25}
	\begin{tabular}{@{}cccccc@{}} 
		%\begin{tabular}{|c|c|c|}
		\hline
		\multirow{2}{4cm}{\textbf{Rewards \& Quality requirements}} & \multicolumn{5}{c}{\textbf{$\tau$ Value}} \\
		\cline{2-6}
		& 5 & 10 & 20 & 40 & 80\\
		
		\hline
		
		$\bs{R} = (50, 30, 15, 20)$, $\bs{Q} = (q_H, 1, 1, 1)$ & 18.5015 & 8.09448 & 5.8064 & 2.2304 & 0.9737\\
		
		\hline

		$\bs{R} = (42, 25, 15, 18)$, $\bs{Q} = (q_H, q_H, 1, 1)$ & 25.6999 & 10.5333 & 6.7059 & 4.9442 & 2.8783\\
		
		\hline

		$\bs{R} = (26, 16, 32, 18)$, $\bs{Q} = (q_H, 1, 1, q_H)$ & 15.0662 & 11.5333 & 3.5229 & 1.5609 & 1.0524\\

		\hline
		
	\end{tabular}
	\vspace{-5pt}
\end{table}

\end{document}